# Spin polarized two-dimensional electron gas embedded in semimagnetic quantum well : ground state, spin responses, spin excitations, Raman spectrum

Florent PEREZ

Institut des Nanosciences de Paris, UMR 7588, CNRS/Université Paris VI

Campus Boucicaut, 140 rue de Lourmel, 75015 PARIS, France

We present theoretical aspects of spin polarized two dimensional electron gas (SP2DEG) which can be achieved in doped semimagnetic quantum wells. This original model system has been recently studied by magneto Raman scattering experiments has given a new access to spin resolved excitations and spectrum of the SP2DEG. Starting from the Diluted Magnetic Semiconductor (DMS) Hamiltonian in presence of the Coulomb interaction between conduction electrons, we define the conditions to reach such a SP2DEG. The equilibrium state is studied at low temperature; in particular a theory for the degree of spin polarization is derived. Dynamical spin susceptibilities are further calculated in the framework of a spin density functional formalism already developed in the past. We then derive spin conserving and spin flip excitations dispersions using a recent determination of the SP2DEG correlation energy corrected from the thickness of the well. The SP2DEG presents two key features: the spin flip wave, which existence is a direct consequence of the Coulomb interaction between the spin polarized electrons, with a dispersion and energy range typical to the SP2DEG obtained in DMS, the spin density fluctuations exhibiting a specific collective behaviour when the spin polarization is increased. The dissipation spectrum through these excitations is studied in detail.

Particular attention is given to the spectrum determined by resonant Raman scattering. We show, indeed, that the latter gives unique access to the spin-fluctuations spectrum of the SP2DEG.

## 1. INTRODUCTION

In the past, high mobility doped semiconductor heterostructures[1] have proven to be a model system for the study of low energy excitations of the two dimensional interacting electron system. Well defined excitations of the Fermi disk have been investigated at very low temperatures by intra-band spectroscopy. Since the energy of these excitations is comparable to the Fermi energy (a few meV), far infrared transmission and electronic resonant Raman scattering (ERRS)[2] in the visible range are the most powerful methods for such a purpose. ERRS has been able to probe excitations of the unpolarized two dimensional electron gas (2DEG) with non-zero in plane momentum where many-body interactions manifest themselves[3,4]. In the integer quantum Hall regime, spin-waves, inter-Landau level magnetoplasmons and spin flip waves (SFW) have been evidenced[5]. Spin excitations of fractional states have also been investigated.[6]

Recently, high mobility spin-polarized two dimensional electron gas (SP2DEG) have been obtained in dilute magnetic semiconductor (DMS) heterostructures like $Cd_{1-x}Mn_xTe/Cd_{1-y}Mg_yTe$ n-type modulation doped quantum wells[7] and investigated by ERRS[8,9,10]. The giant Zeeman effect[11] occurring in these systems allows the creation of a highly spin polarized electron gas (SP2DEG), a model situation in which the spin quantization occurs without direct modification of the orbital motion. This is indeed possible, because magnetic field below 4T are requested, such that when applying the field parallel to the quantum well plane, the Landau orbital quantization remains negligible compared to the well confinement (the magnetic length is always greater than the well-width) and this does not induce significant change in the electron mass[12]. The spin quantization energy, however, can be as high as the Fermi energy, even for usual densities ($\sim 10^{11}$cm$^{-2}$). This provides a novel situation which is exactly the reverse situation to that of GaAs-based systems where Landau quantization dominates over spin quantization. GaAs spin-polarized 2DEG obtained in the past[5,13] were indeed pinned in quantum Hall states. In such situation the Coulomb interaction between electrons is strongly modified by the magnetic field and the excitation spectrum reveals the specific nature of this spin polarized insulating 2D system. We claim that the SP2DEG described here, is conducting, the properties of its excitations carry another richness. Indeed, leaving the spin degeneracy while keeping the electron kinetic energy unperturbed gives unique insight into spin resolved Coulomb interactions[14] and spin responses. These issues have been already extensively addressed from the theoretical point of view[15,16,17,18,19]. But discrimination between relevant theories, accompanying with the intention to fit the experiments is lacking. In this paper, we want first to establish the limits for such an electron gas embedded in a dilute magnetic semiconductor quantum well (with magnetic impurities) to be considered as a model test bed for the SP2DEG. Second, using the formalism developed in Ref. 20 we will derive both the long and transverse spin responses of the SP2DEG. From the response functions, we will determine the dispersions and nature of collective spin excitations. In the third part, we will investigate the dissipation spectrum from the imaginary part of the response functions and we will particularly consider the spectrum determined by Raman scattering measurements[21].

Compared to the amount of work devoted to spin waves in ferromagnetic metals[22,23], collective spin excitations of the paramagnetic electron gas have drawn much less attention in the past. This is partly due to the fact that paramagnetic metals give access to very small spin polarization degree $\zeta \sim 0$. The overall results presented here show how the SP2DEG achieved in DMS quantum wells is an original situation: its physics resembles that of a paramagnetic metal except that the spin polarization degree is here comparable to that of a ferromagnetic metal.

## 2. MODELLING THE SPIN POLARIZED 2D ELECTRON GAS IN A $II_{1-x}Mn_xVI$ QUANTUM WELL

We consider in this section a diluted magnetic modulation doped quantum well where magnetic Mn impurities have been inserted in the well with a fraction $x$ on element II cation sites, e.g. $Cd_{1-x}Mn_xTe$. In such quantum well, two types of systems have to be considered for the physics we are interested in. The first sub-system is composed by two dimensional electrons populating the first confined level of the well. We note $\phi(y)$, the envelope wavefunction of the confined state, $y$-axis being the growth direction. These electrons originate from the n-type dopant impurities located in the barrier. They form an itinerant spin sub-system which is coupled to the second sub-system formed by spins of electrons localized on the manganese impurities introduced in the well. These electrons occupy the d-shell of Mn atoms and each Mn atom will behave like a unique 5/2 spin. A magnetic field will be applied in a direction parallel to the quantum well plane.

### 1.a The DMS Hamiltonian

The coupling Hamiltonian is conventionally written in terms of spin densities, with the Heisenberg convention:

$$\hat{H}_{\mathrm{sd}} = -\alpha \int\int \phi^2(y)\ \tfrac{1}{2}\hat{\mathbf{s}}(\mathbf{r}_{//})\cdot\hat{\mathbf{S}}(\mathbf{r}_{//},y)\,d^2r_{//}dy \tag{1}$$

where $\alpha$ is the s-d exchange integral between s-conduction and d-Mn electrons ($\alpha$>0) [24]. Three dimensional spatial coordinates have been splitted into ($\mathbf{r}_{//}$,$y$), $\mathbf{r}_{//}$ is the $x$-$z$ plane projection parallel to the well. We express $\hat{\mathbf{s}}(\mathbf{r}_{//})$ , the 2D conduction electron spin density:

$$\hat{\mathbf{s}}(\mathbf{r}_{//}) = \sum_{\sigma\sigma'} \hat{\Psi}^{+}_{\sigma}(\mathbf{r}_{//})\,\overline{\boldsymbol{\tau}}_{\sigma\sigma'}\hat{\Psi}_{\sigma'}(\mathbf{r}_{//}) \tag{2}$$

in terms of 2D field operators $\hat{\Psi}^{(+)}_{\sigma}(\mathbf{r}_{//})$ and a vector of Pauli matrices: $\overline{\boldsymbol{\tau}} = (\overline{\tau}_x, \overline{\tau}_y, \overline{\tau}_z)$. For later convenience we also define the 2×2 identity matrix $\overline{\tau}_n$ and the corresponding particle density operator $\hat{n}(\mathbf{r}) = \sum_{\sigma} \hat{\Psi}^{+}_{\sigma}(\mathbf{r})\overline{\tau}_{n,\sigma\sigma}\hat{\Psi}_{\sigma}(\mathbf{r})$. $\hat{\mathbf{S}}(\mathbf{r}_{//},y)$ is the 3D density operator of the 5/2 Mn-electrons spins localized on cation sites $\mathbf{R}_i$ and writes:

$$\hat{\mathbf{S}}(\mathbf{r}_{//},y) = \sum_i \hat{\mathbf{S}}_i \delta(\mathbf{r}-\mathbf{R}_i) \tag{3}$$

A static magnetic field $\mathbf{B}_0$=$B_0\mathbf{z}$ is applied parallely to the plane of the well and we will choose $z$ as the spin quantization axis. With in-plane magnetic field below 4T, the minimum electron magnetic length $l_m = (\hbar/eB_0)^{1/2}$ remains comparable to a typical quantum well width $w$ (150Å). It renders the magnetic orbital quantization negligible. The mass enhancement due to the magneto-hybrid band-bending is also negligible[12]. Hence, the standard DMS Hamiltonian of the two coupled sub-systems in presence of the field is naturally defined by[25]:

$$\hat{H}_{\mathrm{DMS}} = \hat{H}_{gas} + \hat{H}_{sd} + \hat{H}_{\mathrm{Mn}} \tag{4}$$

$\hat{H}_{\mathrm{Mn}}$ is the Mn-spins Hamiltonian:

$$\hat{H}_{\mathrm{Mn}} = g_{Mn}\mu_B \iint \mathbf{B}_0 \cdot \hat{\mathbf{S}} \tag{5}$$

Where $\mu_{\mathrm{B}}$ is the electron Bohr magneton ($\mu_{\mathrm{B}}$>0), $g_{Mn}$ is the Mn electrons g-factor and the direct antiferromagnetic coupling between Mn-spins has been neglected. This coupling results in pairing of Mn-spins[24] which reduces the average amount of spin per cation site $x$ to $x_{eff}$.

$\hat{H}_{gas}$ is the two dimensional electron gas (2DEG) Hamiltonian in the presence of a static magnetic field $\mathbf{B}_0$ having no effect on the kinetic part:

$$\begin{aligned} \hat{H}_{gas} = & \int d^2 r_{//} \left\{ \sum_{\sigma} \hat{\Psi}^{+}_{\sigma}(\mathbf{r}_{//}) \left( -\frac{\hbar^2 \Delta}{2m_b} \right) \hat{\Psi}_{\sigma}(\mathbf{r}_{//}) \right\} + g_e \mu_B \int \mathbf{B}_0 \cdot \tfrac{1}{2} \hat{\mathbf{s}}(\mathbf{r}_{//}) d^2 r_{//} \\ & + \frac{1}{2} \iint d^2 r_{//} d^2 r'_{//} \sum_{\sigma\sigma'} \hat{\Psi}^{+}_{\sigma}(\mathbf{r}_{//}) \hat{\Psi}^{+}_{\sigma'}(\mathbf{r}'_{//}) V(\mathbf{r}_{//} - \mathbf{r}'_{//}) \hat{\Psi}_{\sigma'}(\mathbf{r}'_{//}) \hat{\Psi}_{\sigma}(\mathbf{r}_{//}) \end{aligned} \tag{6}$$

In Eq. (6), $g_e$ and $m_{\mathrm{b}}$ are respectively the conduction electron g-factor and the effective mass, $V(\mathbf{r}_{//} - \mathbf{r}'_{//}) = \frac{e^2}{4\pi\varepsilon_s} \int dy dy' \phi(y) \phi(y') \left[ (\mathbf{r}_{//} - \mathbf{r}'_{//})^2 + (y - y')^2 \right]^{-1/2}$ is the bare Coulomb interaction, $\varepsilon_s$ is the semi-conductor static dielectric constant.

At sufficiently low temperature, 3D DMS systems similar to the one described here, can undergo a ferromagnetic transition[26]. But, in 2D systems with continuous spin-rotational invariance (without spin-orbit interaction) as the one in $\hat{H}_{\rm DMS}$, thermal and quantum spin fluctuations renders long-range magnetic order impossible[27]. Ferromagnetic transition has nevertheless been observed in 2D p-type $Cd_{1-x}Mn_xTe$ quantum wells[28]. This was further demonstrated to be possible because of the specific spin orientation favored by spin-orbit coupling in asymmetric quantum wells[25]. For electron system, spin-orbit coupling remains negligible at usual densities and a ferromagnetic transition has never been observed. For 2DEGs embedded in DMS, the paramagnetic state is the one commonly achieved and we will restrict our discussion on $Cd_{1-x}Mn_xTe$ quantum wells for which experimental observations are available.

### 1.b The paramagnetic spin polarized 2DEG

In the paramagnetic state, the two spin subsystems introduced in Eq. (1) are weakly coupled we can then assume that their spin dynamics behave independently. This is obviously true when the magnetic field is low enough not to bring the Larmor's precession of the two spin systems into resonance[25]. Limitations of this assumption will be clarified later. Therefore, conduction electrons move in a thermalized bath of Mn spins and the conduction spin density couples to the average Mn-spin density. This standard mean-field approximation means that each Mn spin has been frozen in the same statistic thermal average state $\left\langle \mathbf{S}(B_0,T)\right\rangle$ determined by the presence of $\mathbf{B}_0$. $\left\langle \mathbf{S}(B_0,T)\right\rangle$ is given by the modified Brillouin function[24]:

$$\left\langle \mathbf{S}(B_0,T)\right\rangle = \left\langle S_z(B_0,T)\right\rangle \mathbf{z} = \frac{5}{2} B_{5/2}(B_0,T)\,\mathbf{z} \qquad (7)$$

Reciprocally, Mn spins are coupled to a frozen and non-fluctuating conduction spin density. We can then replace the coupling Hamiltonian $\hat{H}_{\mathrm{sd}}$ by two mean-field exchange Hamiltonians:

$$\hat{H}_{sd} \approx -x_{eff}\bar{N}_0\alpha\int \tfrac{1}{2}\hat{\mathbf{s}}(\mathbf{r}_{//})\cdot\langle\mathbf{S}(B_0,T)\rangle d^2r_{//} - \alpha\int\int\phi^2(y)\hat{\mathbf{S}}(\mathbf{r}_{//},y)\cdot\tfrac{1}{2}\langle\mathbf{s}\rangle\, d^2r_{//}dy \tag{8}$$

In the first term, the $y$ integral domain has been cut by the homogeneous Mn distribution in the well of width $w$, therefore, $\bar{N}_0 = N_0\int_0^w \phi^2(y)dy$ where $N_0$ is the number of cation sites per unit volume.

The first and the second term in Eq. (8) lead respectively to the Overhauser and Knight shifts, as they respectively shift the normal Zeeman energies $g_e\mu_B B_0$ and $g_{Mn}\mu_B B_0$ of the conduction and Mn electrons. In the following, as we are interested in describing the SP2DEG, we will concentrate on the conduction spin degrees of freedom by keeping in $\hat{H}_{\mathrm{DMS}}$ only the corresponding sub-system Hamiltonian $\hat{H}_{gas}$ and the Overhauser part of the mean-field Hamiltonian given in Eq.(8). We then deduce the SP2DEG Hamiltonian $\hat{H}$ :

$$\begin{aligned}\hat{H} &= \hat{H}_0 + \hat{H}_{\mathrm{Coul}} = \int d^2r_{//}\left\{\sum_\sigma \hat{\Psi}^+_\sigma(\mathbf{r}_{//})\left(-\frac{\hbar^2\Delta}{2m_b} + \frac{Z(B_0)}{2}\overline{\tau}_{z,\sigma\sigma}\right)\hat{\Psi}_\sigma(\mathbf{r}_{//})\right\} \\ &+\frac{1}{2}\int\int d^2r_{//}d^2r'_{//}\sum_{\sigma\sigma'}\hat{\Psi}^+_\sigma(\mathbf{r}_{//})\hat{\Psi}^+_{\sigma'}(\mathbf{r}'_{//})V(\mathbf{r}_{//}-\mathbf{r}'_{//})\hat{\Psi}_{\sigma'}(\mathbf{r}'_{//})\hat{\Psi}_\sigma(\mathbf{r}_{//})\end{aligned} \tag{9}$$

where $Z(B_0)$ is the total bare Zeeman energy of the conduction electrons, sum of the Overhauser shift (so-called giant Zeeman effect) and the normal Zeeman term:

$$Z(B_0) = -x_{eff}\bar{N}_0\alpha\langle S_z(B_0,T)\rangle + g_e\mu_B B_0 \tag{10}$$

In $Cd_{1-x}Mn_xTe$, the Mn g-factor $g_{Mn}$ equals 2.007, the s-d exchange integral[11] $N_0\alpha$ equals 0.22 eV and the electron g-factor[29] $g_e$ is -1.64. This means that Mn spins tend to align conduction electrons spins parallel to the field, through the s-d exchange, while the normal coupling, has an opposite effect. Hence, in $Z(B_0)$ Overhauser shift and normal

Zeeman have opposite signs. As the former saturates for sufficiently high fields, $Z(B_0)$ reaches a maximum value that depends on temperature and Mn concentration (see Fig. 1).

In our assumption where the dynamical coupling to Mn-spins is negligible, the energy $Z(B_0)$ represents the total external static magnetic action upon the electrons of the well. SP2DEG in $Cd_{1-x}Mn_xTe$ have been realized for typical values $n_{2D}$=2.5 $10^{11}$cm$^{-2}$, $x$~1%, where $\zeta$ was found[8] to be close to 100%.

According to the work of Refs 30 and 25, where the coupled spin dynamics has been theoretically investigated, but in a framework where the Coulomb interaction between carriers was suppressed, we can deduce two criteria for the validity of the decoupling assumption: (1) the ratio $c = n_{2D}\zeta / 2x_{eff} N_0 wS$ between the populations of spin 1/2 and spin $S$=5/2, has to be negligibly small. We find $c$~8.10$^{-5}$ in usual conditions corresponding to an electron sheet density $n_{2D}$=3.10$^{11}$cm$^{-2}$, a spin polarization degree $\zeta = (n_\uparrow - n_\downarrow)/(n_\uparrow + n_\downarrow) = 50\%$, a well width $w$=150Å and Mn concentration $x$~1% (in CdTe, $xN_0$=0.15nm$^{-3}$). (2) Characteristic frequencies of both spin dynamics have to be well separated; this condition is fulfilled if the magnetic field $B_0$ is far from the resonant field $B_R$ where Zeeman energies of conduction and Mn- electrons cross each other[31]:

$$x_{eff}\bar{N}_0\alpha\left|\left\langle S_z\left(B_R,T\right)\right\rangle\right| = \left(\left|g_e\right| + g_{Mn}\right)\mu_B B_R \tag{11}$$

For the above conditions, we find $B_R$~22.5T. If $x$ is 0.2%, $B_R$ reduces to 5T.

By keeping only the mean-field component of $\hat{H}_{sd}$ in $\hat{H}$ , we have dropped magnetic disorder due to random positioning and distributed thermal fluctuations of Mn spins. This allows us to keep the translational invariance symmetry along the quantum well plane. This assumption might be valid until the Fermi wavevector $k_F$ is much smaller than the inverse of the Mn-Mn average distance $\bar{d}_{Mn}$. For the typical values cited above, we find $k_F \bar{d}_{Mn} \approx 0.1$.

It is remarkable that, due to the DMS giant Zeeman effect, the SP2DEG Hamiltonian described by Eq.(9) and (10) is similar to that of a paramagnetic metal except that the spin polarization degree can be comparable to the one of a ferromagnetic metal as we will show in the next section.

### 1.c SP2DEG equilibrium state

As we have spin-rotational invariance along the spin-quantization axis, the equilibrium state of this model system is totally characterized by its equilibrium spin polarization degree $\zeta$, its density $n_{2D}$ and the temperature $T$ which will be taken equal to 0K in this section. It is already well established that exchange and correlation Coulomb interactions present in $H_{\text{Coul}}$ enhance the spin susceptibility $\chi$ of an electron gas over that of the Pauli spin susceptibility $\chi_0$ for non-interacting electrons[32]. The spin susceptibility enhancement is exactly the inverse of the spin stiffness[14]:

$$\frac{\chi}{\chi_0} = \left( 1 + \frac{r_s^2}{2} \frac{\partial^2 \varepsilon_{xc}}{\partial \zeta^2} \right)^{-1} \qquad (12)$$

where $r_s = \left( a_B^* \sqrt{\pi n_{2D}} \right)^{-1}$ is the ratio of the mean spacing between electrons to the Bohr radius, $a_B^*$ and $\varepsilon_{xc}$ is the exchange-correlation part of the SP2DEG ground-state energy per particle[14], expressed in Rydbergs[33]. In Eq. (12), the spin stiffness has been separated into the kinetic and the exchange-correlation contributions. For $\zeta$=0, it becomes[14]:

$$\frac{\chi}{\chi_0} = \left( 1 - \frac{\sqrt{2}}{\pi} r_s + \frac{r_s^2}{2} \frac{\partial^2 \varepsilon_c}{\partial \zeta^2} \right)^{-1} \qquad (13)$$

where we have further separated exchange and correlation contributions. It is clear that the exchange-correlation part in the spin-stiffness coefficient $\partial^2\varepsilon_{xc}/\partial\zeta^2$ which rules the spin-susceptibility enhancement, has a dominant negative contribution arising from exchange. If correlations were switched off, exchange would make the 2DEG undergo a spontaneous transition to the full polarized state at a critical $r_s = \pi/\sqrt{2} \approx 2.22$. But the positive contribution from correlations drastically reduces the enhancement and shifts the critical $r_s$ to ~26.95, where the 2DEG has been predicted to recover the ferromagnetic state[34]. For usual $r_s$, exchange-correlations make the SP2DEG evolves from the non-interacting ground state with spin polarization $\zeta_0$ given by:

$$\zeta_0 = -m_b Z\left(B_0\right)/2\pi\hbar^2 n_{2D} \tag{14}$$

to the interacting ground state having an enhanced spin polarization degree $\zeta$.

Since $\chi = \partial m_z/\partial b_z$ is the variation of the 2DEG magnetization $m_z \propto n_{2D}\zeta$ with any magnetic field $b_z$ acting on the 2DEG, the spin susceptibility enhancement is linked to the spin polarization enhancement by:

$$\chi/\chi_0 = d\zeta/d\zeta_0 \tag{15}$$

Integration of Eq.(12), combined with Eq. (15), yields the following exact result for the spin polarization enhancement:

$$\frac{\zeta}{\zeta_0} = \left(1 + \frac{r_s^2}{2}\frac{1}{\zeta}\frac{\partial\varepsilon_{xc}}{\partial\zeta}\right)^{-1} \tag{16}$$

The spin polarization degree $\zeta$ is basically an ensemble property of the SP2DEG and is, contrary to $\zeta_0$ a measurable quantity.

In Fig. 2, the maximum achievable value for the spin polarization degree has been plotted as a function of $n_{2D}$ and of the nominal concentration $x$ of Mn for a typical $Cd_{1-x}Mn_xTe$/CdMgTe quantum well of width $w$=150Å. (Fig. 2(a)) and (Fig. 2(b)) compares respectively the calculation without ($\zeta_0$) and with ($\zeta$) interactions. The maximum Zeeman energy is the maximum of Eq. (10), calculated for T=1.5K. The Exchange-correlation energy has been taken from Ref. 14 (T=0K calculation), corrected by finite thickness[36] effects. The concentration of Mn has been kept low (3%) to avoid structural disorder neglected in the mean field approximation of Eq. (8). For the lowest densities, we see the interaction enhancement of the spin polarization degree which renders the full spin polarized state achievable for Mn and electron concentration respectively around 1% and 2.0 $10^{11}$cm$^{-2}$.

It is convenient to link $\zeta$ to single particle properties in a manner like the non-interacting spin-polarization degree $\zeta_0$ is linked to the bare mass $m_b$ and the bare Zeeman energy $Z$ in Eq. (14). The enhancement found in Eq. (16) suggests introducing a renormalized mass $m^*$ and a renormalized Zeeman energy $Z^*$, to obtain an expression of $\zeta$ equivalent to Eq. (14), but valid for the interacting case:

$$\zeta = -m^*(B_0)Z^*(B_0)/2\pi\hbar^2 n_{2D} \tag{17}$$

Eqs (14) and (17) give the relation between renormalized and non-interacting quantities:

$$\frac{\zeta}{\zeta_0} = \frac{m^* Z^*}{m_b Z} \tag{18}$$

Despite their apparent mathematical definition $m^*$ and $Z^*$ are usual Fermi liquid parameters and their determination in 2DEG has drawn strong experimental and theoretical interest. Further separation of the mass enhancement $m^*/m_b$ from the Zeeman enhancement $Z^*/Z$ in Eq. (18) requires the derivation of a spin-resolved self energy for the spin-polarized 2DEG. The self-energy is in itself a cumbersome problem still unsolved. An accurate determination of the self-energy and the renormalized mass valid for an unpolarized 2DEG has been done recently[35]. The evaluation took into account corrections due to the exchange-correlation ground state energy parameterized after quantum Monte Carlo calculations[14], corrected from the thickness of the well[36]. It was found that for intermediate $r_s$ values and typical confinement lengths around 150Å, the mass correction is negative and is about -5%. This was confirmed by magneto-transport measurements[37]. In the spin-polarized case, strong modifications of the spin resolved self-energy have been predicted in the random phase approximation (RPA)[38]. But RPA is known to give very approximate corrections for $r_s$ above unity where correlations play an important role. Indeed, the spin dependent masses $m^*_{\uparrow,\downarrow}$ and their dependence with $\zeta$, calculated in Ref. 38, showed a poor agreement with measurements in Ref. 8 carried out at $r_s$~2.5. More quantitatively, RPA calculations predict very strong non linear enhancement of the spin dependent masses when the SP2DEG reaches the full polarized state, even for values of $r_s$ as low as 2. Its origin is in the divergence of the second derivative of the exchange energy. As described above and in Ref. 14, correlations cancel out this divergence, at least for usual values of $r_s$. The mass enhancement due to spin polarization is certainly much less than the RPA prediction of Ref. 38. For $r_s$=2 and $\zeta$=-50%, the latter gives: $m^*_\downarrow(\zeta=-0.5)/m^*_\downarrow(\zeta=0)\sim 6\%$. Adding the negative mass enhancement of Ref. 35, we conclude that these theories are not sufficient to determine the sign of the mass correction, but its modulus is probably close to 2-3%. For the same conditions the spin polarization enhancement given by Eq. (16) is 85%. We then deduce, from Eq. (18), that renormalization of the Zeeman energy captures most of the enhancement. In his pioneering developments[20], A.K. Rajagopal proposed to approximate the self-energy, which is a ground-state property, by spin-density functional potentials introduced in the spin resolved Kohn-Sham equations[39]. These equations give eigenvalues and orbitals which lead to the correct many-body ground-state energy and equilibrium densities. Differences between Kohn-Sham eigenvalues are also known to be good approximations of quasi-particle excitations energies[40]. We show in the next section that this approximation leads to a determination of $Z^*$ which captures the entire enhancement, but is actually the more satisfying calculation. Moreover, the self energy expressed as a functional of densities is compatible with the derivation of response functions, which will be developed in section 3.

### 1.d Approximated self-energy

To define the self-energy, we must introduce the spin resolved Green's functions:

$$\bar{G}_{\sigma\sigma'}(\mathbf{r}t,\mathbf{r}'t') = -i\left\langle T\left[\hat{\Psi}_{\sigma}(\mathbf{r}t)\hat{\Psi}^{+}_{\sigma'}(\mathbf{r}'t')\right]\right\rangle \quad (19)$$

Where $\mathbf{r}_{//}$ has been replaced by $\mathbf{r}$ for simplicity, $T[\ ]$ is the time ordering operator , $\hat{\Psi}^{(+)}_{\sigma}(\mathbf{r}t)$ are Heisenberg operators of $\hat{\Psi}^{(+)}_{\sigma}(\mathbf{r})$, whose time evolution is determined by the SP2DEG Hamiltonian $\hat{H}$ , and $\langle ... \rangle$ stands for the grand canonical ensemble average. The 2×2 Green function matrix $\bar{G}(\mathbf{r}t,\mathbf{r}'t')$ obeys the Dyson equation:

$$\left[i\hbar\frac{\partial}{\partial t}\bar{\tau}_n + \frac{\hbar^2\Delta}{2m_b}\bar{\tau}_n - \frac{Z}{2}\bar{\tau}_z\right]\bar{G}(\mathbf{r}t,\mathbf{r}'t') - \int dt''\int d^2r''\bar{\Sigma}(\mathbf{r}t,\mathbf{r}''t'')\bar{G}(\mathbf{r}''t'',\mathbf{r}t') = \hbar\bar{\tau}_n\delta(\mathbf{r}t-\mathbf{r}'t') \quad (20)$$

where $\bar{\Sigma}$ is the self-energy and $\delta$ is the three dimensional delta-function. The spin-density functional approximation of the self-energy writes[20]:

$$\overline{\Sigma}(\mathbf{r}t,\mathbf{r}'t') \approx \left\{\overline{\tau}_n V_H[\mathbf{r}t,n] + \overline{\tau}_n V_{xc}[\mathbf{r}t,n,\mathbf{s}] + \overline{\boldsymbol{\tau}}\cdot\mathbf{W}_{xc}[\mathbf{r}t,n,\mathbf{s}]\right\}\delta(\mathbf{r}t-\mathbf{r}'t') \quad (21)$$

where $n$ and $\mathbf{s}=(s_x,s_y,s_z)$ have to be understood as densities taken at the position $\mathbf{r}t$, and we have introduced the Kohn-Sham potentials:

$$V_H[\mathbf{r}t,n] = \frac{e^2}{4\pi\varepsilon_s}\iint dy dy' \phi(y)\phi(y') \int \frac{n(\mathbf{r}'t)}{\sqrt{(\mathbf{r}-\mathbf{r}')^2+(y-y')^2}} d^2r' \quad (22)$$

$$\begin{aligned} V_{xc}[\mathbf{r}t,n,\mathbf{s}] &= \partial E_{xc}[n,\mathbf{s}]/\partial n(\mathbf{r}t) \\ \mathbf{W}_{xc}[\mathbf{r}t,n,\mathbf{s}] &= \partial E_{xc}[n,\mathbf{s}]/\partial \mathbf{s}(\mathbf{r}t) \end{aligned} \quad (23)$$

Eq. (23) is the Hartree potential. Eq (24) gives potentials due to the exchange-correlation field. $E_{xc}$ is the ground state energy per unit surface: $E_{xc} = n_{2D} R_y^* \varepsilon_{xc}$, where $R_y^*$ is the effective Rydberg[33]. The spin density approximation of the self-energy is local in space and time as it acts on single particle wave functions in Kohn-Sham Schrödinger-like equations. It includes however a non-local contribution which relies in the density dependence which accounts for the collective system. Therefore, Kohn-Sham single particle wave functions are by no means true single particle states of the many-body system, and the self-energy of Eq. (21) has no guarantee to be a good approximated form. Nevertheless, the differences in single-particle energies given by Kohn-Sham equations are known to be good approximations of single particle excitations energies[40]. At the end, the error committed under these assumptions will be estimated by comparing the spin-polarization degree deduced from single-particle energies and the one obtained from the exact relation of Eq. (16).

For the equilibrium state of the homogeneous SP2DEG, the above densities are position and time independent, the spin-rotational invariance cancels the spin-transverse components, hence $n = n_{2D}$, $\mathbf{s} = \left(0, 0, n_{2D}\zeta\right)$, $\mathbf{W}_{xc}\left[\mathbf{r}t, n, \mathbf{s}\right] = n_{2D}^{-1}\,\partial E_{xc}\left[n,\zeta\right]/\partial\zeta\left(0,0,1\right)$, and the Hartree contribution is cancelled by the positive charge background.

As the SP2DEG Hamiltonian is translationally invariant in space and time, the Fourier transforms of the Green function and the self energy can be defined as:

$$\tilde{\bar{G}}\left(\mathbf{k},\omega\right) = \int dt \int d^2 r \bar{G}\left(\mathbf{r}t, 0\right) e^{-i\mathbf{k}\cdot\mathbf{r}+i\omega t} \qquad (24)$$

Because of the absence of transverse spin components, the self-energy and Green function are diagonal and the solution of the Dyson equation (20) thus becomes:

$$\tilde{\bar{G}}_{\sigma\sigma'}\left(\mathbf{k},\omega\right) = \left(\omega - \varepsilon_{k\sigma}/\hbar - \tilde{\bar{\Sigma}}_{\sigma\sigma}\left(\mathbf{k},\omega\right)/\hbar\right)^{-1}\delta_{\sigma\sigma'} \qquad (25)$$

where we have introduced non-interacting single electron energies: $\varepsilon_{\mathbf{k}\sigma} = \hbar^2 k^2/2m_b + \sigma Z/2$. Quasi-particle energies $\varepsilon^*_{\mathbf{k}\sigma}$ are naturally deduced from the standard equation:

$$\varepsilon^*_{\mathbf{k}\sigma} = \varepsilon_{\mathbf{k}\sigma} + \tilde{\bar{\Sigma}}_{\sigma\sigma}\left(\mathbf{k}, \varepsilon^*_{\mathbf{k}\sigma}\right) \qquad (26)$$

which, using the approximated self-energy of Eq. (21), constant in Fourier space, has a trivial solution. We find, by combining equations (21), (23) and (26):

$$\varepsilon^*_{\mathbf{k}\sigma} = \varepsilon_{\mathbf{k}\sigma} + \partial E_{xc}/\partial n_{2D} + \sigma\, n_{2D}^{-1}\partial E_{xc}/\partial\zeta \qquad (27)$$

In Eq. (27), the k-independent self-energy gives no-mass correction, but the enhanced Zeeman energy is derived:

$$Z^*\left(r_s,\zeta\right) = Z + 2n_{2D}^{-1}\frac{\partial E_{xc}}{\partial\zeta} \qquad (28)$$

Using Eq. (18) and $m^*$=$m_b$, the latter can also be expressed as:

$$Z^*\left(r_s,\zeta\right)/Z = \left(1 + \frac{r_s^2}{2}\frac{1}{\zeta}\frac{\partial\varepsilon_{xc}}{\partial\zeta}\right)^{-1} \qquad (29)$$

After a comparison between Eq. (29) and Eq. (18), it becomes evident that the renormalized Zeeman energy found using the spin density functional version of the self-energy, has captured the mass renormalization. As stated above, we expect to get an error of a few percent contained in the renormalized $Z^*$ of Eq. (30). We remind that this error arising from a bad separation between the mass and Zeeman renormalization contributions does not influence the spin-polarization degree which remains a reliable quantity.

## 3. SPIN RESOLVED RESPONSE FUNCTIONS OF THE SP2DEG

In this section, the dynamical response of the SP2DEG will be evaluated by perturbing the Hamiltonian of Eq. (9) with an external electric field $\varphi(\mathbf{r},t)=\tilde{\varphi}(q,\omega)e^{i\mathbf{q}\cdot\mathbf{r}-i\omega t}$ and a magnetic field $\mathbf{b}(\mathbf{r},t)=\tilde{\mathbf{b}}(q,\omega)e^{i\mathbf{q}\cdot\mathbf{r}-i\omega t}$, both varying sinusoidally in time and space with a frequency $\omega$ and a 2-dimensional wave vector $\mathbf{q}$ parallel to the quantum well. The resulting perturbing Hamiltonian is:

$$\hat{H}_{pert}=\int d^2r\left\{-e\varphi(\mathbf{r},t)\cdot\hat{n}(\mathbf{r})+g_e\mu_B\mathbf{b}(\mathbf{r},t)\cdot\tfrac{1}{2}\hat{\mathbf{s}}(\mathbf{r})\right\} \tag{30}$$

where $\hat{n}(\mathbf{r})=\sum_\sigma\hat{\Psi}^+_\sigma(\mathbf{r})\hat{\Psi}_\sigma(\mathbf{r})$ is the particle density operator and $\hat{\mathbf{s}}(\mathbf{r})$ are spin densities operators defined in Eq. (2). We re-write the perturbing Hamiltonian in a more convenient way:

$$\hat{H}_{pert}=\sum_\alpha\int d^2r\left\{F_\alpha(\mathbf{r},t)\cdot\hat{n}_\alpha(\mathbf{r})\right\} \tag{31}$$

with index $\alpha=n,z,+,-$; $[\hat{n}_\alpha]=(\hat{n},\hat{s}_z,\hat{s}_+,\hat{s}_-)$ and $[F_\alpha]=\left(-e\varphi,\tfrac{1}{2}g_e\mu_Bb_z,\tfrac{1}{2}g_e\mu_Bb_-,\tfrac{1}{2}g_e\mu_Bb_+\right)$.

The transverse operators $\hat{s}_\pm=\hat{s}_x\pm i\hat{s}_y$ and the rotating fields $b_\pm=\left(b_x\mp ib_y\right)/2$ have been introduced.

The set of perturbing fields $[F_\alpha]$ will induce changes in the densities that will be determined in the linear approximation. We will therefore concentrate on the evaluation of the density change having the same Fourier components:

$$\delta\tilde{n}_\alpha(\mathbf{q},\omega)=\iint\left[\left\langle\hat{n}_\alpha(\mathbf{r}t)_F\right\rangle-\left\langle\hat{n}_\alpha(\mathbf{r}t)\right\rangle\right]e^{-i\mathbf{q}\cdot\mathbf{r}+i\omega t}d^2r\,dt \qquad (32)$$

where $\langle\ldots\rangle$ denotes the average in the thermal equilibrium ensemble, which in the following will be reduced to the ground-state expectation value as we work at zero temperature. $\hat{n}_\alpha(\mathbf{r}t)_F$ [ $\hat{n}_\alpha(\mathbf{r}t)$ ] is a Heisenberg operator whose time evolution is governed by the SP2DEG Hamiltonian $\hat{H}$ in presence (resp. absence) of the perturbing fields. As the unperturbed Hamiltonian is time independent and translationnally invariant, $\left\langle\hat{n}_\alpha(\mathbf{r}t)\right\rangle$ are constants of space and time and equal ground state equilibrium densities $[n_\alpha]=(n_{2D},n_{2D}\zeta,0,0)$. The linear response functions are defined as:

$$\chi_{\alpha\beta}(\mathbf{q},\omega)=\delta\tilde{n}_\alpha(\mathbf{q},\omega)/\tilde{F}_\beta(\mathbf{q},\omega) \qquad (33)$$

and can also be expressed as follows:

$$\chi_{\alpha\beta}(\mathbf{q},\omega)=-\frac{i}{\hbar}\int_0^\infty dt\int d^2r\left\langle\left[\hat{n}_\alpha(\mathbf{r}t),\hat{n}_\beta\right]\right\rangle e^{-i\mathbf{q}\cdot\mathbf{r}+i\omega t} \qquad (34)$$

### 1.e Core equation of the linear responses

To evaluate $\chi_{\alpha\beta}$, we follow the scheme of A.K. Rajagopal[20]. We will make use of the spin resolved Green's functions in the presence of perturbing fields:

$$\bar{G}_{F,\sigma\sigma'}(\mathbf{r}t,\mathbf{r}'t')=-i\left\langle T\left[\hat{\Psi}_\sigma(\mathbf{r}t)_F\hat{\Psi}^+_{\sigma'}(\mathbf{r}'t')_F\right]\right\rangle \qquad (35)$$

The 2×2 Green function matrix $\bar{G}_F(\mathbf{r}t,\mathbf{r}'t')$ obeys the Dyson equation:

$$\left[i\hbar\frac{\partial}{\partial t}\bar{\tau}_n+\frac{\hbar^2\Delta}{2m_b}\bar{\tau}_n-\frac{Z}{2}\bar{\tau}_z-\sum_\alpha F_\alpha\bar{\tau}_\alpha\right]\bar{G}_F\left(\mathbf{r}t,\mathbf{r}'t'\right)-\int dt''\int d^2r''\bar{\Sigma}_F\left(\mathbf{r}t,\mathbf{r}''t''\right)\bar{G}_F\left(\mathbf{r}''t'',\mathbf{r}'t'\right)=\hbar\bar{\tau}_n\delta\left(\mathbf{r}t-\mathbf{r}'t'\right)$$
(36)

where $\bar{\Sigma}_F$ is the self-energy disturbed by density changes induced by the fields. Given the expressions of densities:

$$\left\langle\hat{n}_\alpha\left(\mathbf{r}t\right)_F\right\rangle=-iTr\left[\bar{\tau}_\alpha\bar{G}_F\left(\mathbf{r}t,\mathbf{r}t^+\right)\right]\qquad\left\langle\hat{n}_\alpha\left(\mathbf{r}t\right)\right\rangle=-iTr\left[\bar{\tau}_\alpha\bar{G}\left(\mathbf{r}t,\mathbf{r}t^+\right)\right]\tag{37}$$

the linear response functions can also be written (for infinitesimally small perturbations) as:

$$\chi_{\alpha\beta}\left(\mathbf{q},\omega\right)=-i\iint Tr\left[\bar{\tau}_\alpha\frac{\partial\bar{G}_F}{\partial\tilde{F}_\beta}\left(\mathbf{r}t,\mathbf{r}t^+\right)\right]e^{-i\mathbf{q}\cdot\mathbf{r}+i\omega t}d^2rdt\tag{38}$$

A.K. Rajagopal has found a convenient expression of the first order contribution to $\frac{\partial\bar{G}_F}{\partial\tilde{F}_\beta}\left(\mathbf{r}t,\mathbf{r}t^+\right)$ in terms of unperturbed Green's functions, it writes:

$$\frac{\partial\bar{G}_F}{\partial\tilde{F}_\beta}\left(\mathbf{r}t,\mathbf{r}t^+\right)=-\int\int\bar{G}\left(\mathbf{r}t,\mathbf{u}\right)\frac{\partial\bar{G}_F^{-1}}{\partial\tilde{F}_\beta}\left(\mathbf{u},\mathbf{v}\right)\bar{G}\left(\mathbf{v},\mathbf{r}t^+\right)d\mathbf{u}d\mathbf{v}\tag{39}$$

where the following property of the inverse Green's function matrix $\bar{G}_F^{-1}\left(\mathbf{r}t,\mathbf{r}'t'\right)$ have been used:

$$\int\bar{G}^{-1}\left(\mathbf{u},\mathbf{v}\right)\bar{G}\left(\mathbf{v},\mathbf{w}\right)d\mathbf{v}=\int\bar{G}\left(\mathbf{u},\mathbf{v}\right)\bar{G}^{-1}\left(\mathbf{v},\mathbf{w}\right)d\mathbf{v}=\bar{\tau}_0\delta\left(\mathbf{u}-\mathbf{w}\right)\tag{40}$$

The Dyson equation (36) leads to $\bar{G}_F^{-1}$:

$$\bar{G}_F^{-1}\left(\mathbf{r}t,\mathbf{r}'t'\right)=\hbar^{-1}\left[\left(i\hbar\frac{\partial}{\partial t}+\frac{\hbar^2\Delta}{2m_b}\right)\bar{\tau}_n-\frac{Z}{2}\bar{\tau}_z-\sum_\alpha F_\alpha\left(\mathbf{r},t\right)\bar{\tau}_\alpha\right]\delta\left(\mathbf{r}t-\mathbf{r}'t'\right)-\hbar^{-1}\bar{\Sigma}_F\left(\mathbf{r}t,\mathbf{r}'t'\right)\tag{41}$$

and its derivatives:

$$\frac{\partial\bar{G}_F^{-1}}{\partial\tilde{F}_\beta}\left(\mathbf{r}t,\mathbf{r}'t'\right)=\hbar^{-1}\bar{\tau}_\alpha e^{i\mathbf{q}\cdot\mathbf{r}-i\omega t}\delta\left(\mathbf{r}t-\mathbf{r}'t'\right)-\hbar^{-1}\frac{\partial\bar{\Sigma}_F}{\partial\tilde{F}_\beta}\left(\mathbf{r}t,\mathbf{r}'t'\right)\tag{42}$$

Insertion of expression (42) into (39) and (38) yields:

$$\chi_{\alpha\beta}(\mathbf{q},\omega)=\chi_{\alpha\beta}^{(0)}(\mathbf{q},\omega)-\frac{i}{\hbar}\iint d^2rdt\cdot e^{-i\mathbf{q}\cdot\mathbf{r}+i\omega t}\iint Tr\left[\overline{\tau}_\alpha\overline{G}(\mathbf{r}t,\mathbf{u})\frac{\partial\overline{\Sigma}_F}{\partial\tilde{\overline{F}}_\beta}(\mathbf{u},\mathbf{v})\overline{G}(\mathbf{v},\mathbf{r}t^+)\right]d\mathbf{u}d\mathbf{v} \quad (43)$$

where

$$\chi_{\alpha\beta}^{(0)}(\mathbf{q},\omega)=-\frac{i}{\hbar}\iiiint Tr\left[\overline{\tau}_\alpha\overline{G}(\mathbf{r}t,\mathbf{r}'t')\overline{\tau}_\beta\overline{G}(\mathbf{r}'t',\mathbf{r}t^+)\right]e^{-i\mathbf{q}\cdot(\mathbf{r}-\mathbf{r})+i\omega(t-t')}d^2rd^2r'dtdt' \quad (44)$$

are response functions describing the densities responses in absence of dynamical screening due to the perturbation *F*. Using the Fourier transform of the unperturbed Green's functions, expression (44) can be simplified into:

$$\chi_{\alpha\beta}^{(0)}(\mathbf{q},\omega)=-\frac{i}{\hbar}\iint Tr\left[\overline{\tau}_\alpha\tilde{\overline{G}}(\mathbf{k}+\mathbf{q},\varepsilon+\omega)\overline{\tau}_\beta\tilde{\overline{G}}(\mathbf{k},\varepsilon)\right]\frac{d^2k}{(2\pi)^2}\frac{d\varepsilon}{2\pi} \quad (45)$$

which are combinations of the spin resolved Lindhard-type polarizability:

$$\begin{aligned}\Pi_{\sigma\sigma'}(\mathbf{q},\omega)&=-\frac{i}{\hbar}\iint\tilde{\overline{G}}_{\sigma\sigma}(\mathbf{k}+\mathbf{q},\varepsilon+\omega)\tilde{\overline{G}}_{\sigma'\sigma'}(\mathbf{k},\varepsilon)\frac{d\varepsilon}{2\pi}\\&=\int\frac{d^2k}{(2\pi)^2}\frac{n_{\mathbf{k}\sigma}-n_{\mathbf{k}+\mathbf{q}\sigma'}}{\hbar\omega+\varepsilon_{\mathbf{k}\sigma}^*-\varepsilon_{\mathbf{k}+\mathbf{q}\sigma'}^*+i\hbar\eta}\end{aligned} \quad (46)$$

In Eq. (46), $n_{\mathbf{k}\sigma}$ are state occupation numbers. We have explicitly introduced the frequency imaginary part stating for homogeneous broadening. Because of spin-rotational invariance along the magnetization axis, there is an important reduction of non-zero $\chi_{\alpha\beta}^{(0)}(\mathbf{q},\omega)$. Only the following elements remain:

$$\chi_{nn}^{(0)}=\chi_{zz}^{(0)}=\Pi_{\uparrow\uparrow}+\Pi_{\downarrow\downarrow},\ \chi_{nz}^{(0)}=\chi_{zn}^{(0)}=\Pi_{\uparrow\uparrow}-\Pi_{\downarrow\downarrow},\ \chi_{+-}^{(0)}=4\Pi_{\uparrow\downarrow},\ \chi_{-+}^{(0)}=4\Pi_{\downarrow\uparrow} \quad (47)$$

Further derivation of the core Eq. (43) requires additional assumptions for the self-energy in the presence of the perturbation.

### 1.f Adiabatic spin density approximation

The adiabatic density approximation has proven to be very successful in the derivation of response functions leading to collective excitation frequencies. The adiabatic approximation assumes that the perturbed self-energy is equal to the evaluation of the unperturbed self energy functional, taken at the instantaneous densities $\left\langle \hat{n}_\alpha(\mathbf{r}t)_F \right\rangle$. By doing so, we naturally assume that the external perturbing fields have been switched on and are varying in a time scale much slower than the characteristic time scale of the unperturbed self-energy, the latter corresponding to intrinsic time-scales of the SP2DEG. In fact, it has been demonstrated that the Kohn-Sham self-energy of Eq. (21) is valid over a frequency range limited by correlated multi-particle excitations frequencies[40], which are much higher than the energies of the processes we are interested in. Hence, the variation time scale of the external fields is sufficiently low to assume that the density change governing the self-energy change is the one given by response-functions themselves. To first order in density, the perturbed self-energy is expressed as:

$$\bar{\Sigma}_F(\mathbf{r}t,\mathbf{r}'t') \cong \bar{\Sigma}(\mathbf{r}t,\mathbf{r}'t') + \sum_{\eta,\beta}\left(\frac{\partial\bar{\Sigma}}{\partial n_\eta}\right)\chi_{\eta\beta}(\mathbf{q},\omega)\tilde{F}_\beta e^{i\mathbf{q}\cdot\mathbf{r}-i\omega t} \tag{48}$$

where both $\bar{\Sigma}$ and $\partial\bar{\Sigma}/\partial n_\eta$ are evaluated at the equilibrium densities.

By inserting (48) into the core equation (43), we find the self-consistent equations for the response functions:

$$\chi_{\alpha\beta}(\mathbf{q},\omega) = \chi^{(0)}_{\alpha\beta}(\mathbf{q},\omega) + \sum_{\eta,\upsilon}\chi^{(0)}_{\alpha\eta}(\mathbf{q},\omega)\frac{\partial\tilde{\Sigma}_\eta}{\partial n_\upsilon}(\mathbf{q},\omega)\chi_{\upsilon\beta}(\mathbf{q},\omega) \tag{49}$$

where $\tilde{\Sigma}_\eta$ is the Fourier transform of the potential multiplying $\bar{\tau}_\eta$ in the self-energy (21).

As the exchange-correlation energy in the ground state is a functional of the density and the magnetization amplitude only, we find after some basic algebra the non-zero derivatives of the unperturbed self-energy:

$$\begin{aligned}
&\frac{\partial\tilde{\Sigma}_0}{\partial n} = V_{\mathbf{q}} + G_{nn}^{xc} = \frac{F(q)}{L^2}\frac{e^2}{2\varepsilon_s q} + \frac{\partial^2 E_{xc}}{\partial n_{2D}^2} \\
&\frac{\partial\tilde{\Sigma}_z}{\partial n} = \frac{\partial\tilde{\Sigma}_0}{\partial s_z} = G_{nz}^{xc} = \frac{1}{n_{2D}}\frac{\partial^2 E_{xc}}{\partial n_{2D}\partial\zeta} \\
&\frac{\partial\tilde{\Sigma}_z}{\partial s_z} = G_{zz}^{xc} = \frac{1}{n_{2D}^2}\frac{\partial^2 E_{xc}}{\partial\zeta^2} \\
&\frac{\partial\tilde{\Sigma}_+}{\partial s_-} = \frac{\partial\tilde{\Sigma}_-}{\partial s_+} = G_{+-}^{xc} = \frac{1}{2n_{2D}^2\zeta}\frac{\partial E_{xc}}{\partial\zeta}
\end{aligned} \tag{50}$$

In the first derivative, we have introduced $V_{\mathbf{q}} = F(q)e^2/2\varepsilon_s qL^2$, the space Fourier transform of the bare Coulomb potential, product of the 2D Coulomb interaction with a form factor $F(q)$ that depends on $\phi(y)$ [41]. $L^2$ is the sample area.

The spin-rotational invariance decouples the longitudinal and transverse responses. The former involves single particle excitations that conserve the electron spin, while the latter is composed by spin-flip excitations.

### 1.g Longitudinal response

When solving Eq. (49) for $(\alpha,\beta)\in(0,z)$, we find three coupled equations from which we determine the longitudinal response characterized by coupled charge density and spin density fluctuations excited by the longitudinal perturbing fields:

$$\begin{pmatrix}\delta\tilde{n} \\ \delta\tilde{s}_z\end{pmatrix} = \begin{pmatrix}\chi_{nn} & \chi_{nz} \\ \chi_{zn} & \chi_{zz}\end{pmatrix}\begin{pmatrix}-e\tilde{\varphi} \\ \frac{1}{2}g_e\mu_B\tilde{b}_z\end{pmatrix} \tag{51}$$

with:

$$\begin{aligned}
\chi_{nn} &= \frac{1}{D}\left[\Pi_{\uparrow\uparrow} + \Pi_{\downarrow\downarrow} - 4G_{zz}^{xc}\Pi_{\downarrow\downarrow}\Pi_{\uparrow\uparrow}\right] \\
\chi_{zz} &= \frac{1}{D}\left[\Pi_{\uparrow\uparrow} + \Pi_{\downarrow\downarrow} - 4\left(V_{\mathbf{q}} + G_{nn}^{xc}\right)\Pi_{\downarrow\downarrow}\Pi_{\uparrow\uparrow}\right] \\
\chi_{nz} = \chi_{zn} &= \frac{1}{D}\left[\Pi_{\uparrow\uparrow} - \Pi_{\downarrow\downarrow} + 4G_{nz}^{xc}\Pi_{\downarrow\downarrow}\Pi_{\uparrow\uparrow}\right]
\end{aligned} \tag{52}$$

$$D = 1-\left(V_{\mathbf{q}}+G_{nn}^{xc}+G_{zz}^{xc}\right)\left(\Pi_{\uparrow\uparrow}+\Pi_{\downarrow\downarrow}\right)-2G_{nz}^{xc}\left(\Pi_{\uparrow\uparrow}-\Pi_{\downarrow\downarrow}\right)+4\left[\left(V_{\mathbf{q}}+G_{nz}^{xc}\right)G_{zz}^{xc}-\left(G_{nz}^{xc}\right)^2\right]\Pi_{\downarrow\downarrow}\Pi_{\uparrow\uparrow} \quad (53)$$

Spin conserving excitations appear as poles of the above matrix determinant. They originate from poles of the Lindhard polarizabilities, or from zeros of the denominator *D*. The former are called spin-conserving single particle excitations as they correspond to the change in the kinetic energy of a single electron excited from an occupied state $\left|\mathbf{k},\sigma\right\rangle$ to an empty state $\left|\mathbf{k}+\mathbf{q},\sigma\right\rangle$. Zeros of *D* are the collective excitations. A feature of the SP2DEG is that they are mixed excitations of both the charge and spin densities. These excitations do not induce any change in the spin polarization degree as they do not modify spin up and spin down populations. But each population might acquire a disturbance oscillating in space and time which is in phase in the case of the charge mode and out of phase in the case of the spin density mode.

In the unpolarized limit $\zeta$=0, we have $\Pi_{\uparrow\uparrow}=\Pi_{\downarrow\downarrow}=\frac{1}{2}\Pi$ and $G_{nz}^{xc}=0$. The above susceptibilities simplify in the well-known expressions[42]:

$$\chi_{nn}=\frac{\Pi}{1-\left(V_{\mathbf{q}}+G_{nn}^{xc}\right)\Pi},\ \chi_{zz}=\frac{\Pi}{1-G_{zz}^{xc}\Pi},\ \chi_{nz}=\chi_{zn}=0 \quad (54)$$

Spin and charge density responses are no longer coupled, collective excitations of $\chi_{nn}$ are the pure plasmon branch, while collective excitations of $\chi_{zz}$ are pure spin-density excitations.

### 1.h Transverse response

If we now solve Eq. (49) for $(\alpha,\beta)\in(+,-)$, we find two decoupled equations leading to the transverse response characterized by spin-flip excitations induced by a magnetic field rotating in the plane perpendicular to the magnetization axis:

$$\begin{pmatrix} \delta\tilde{s}_+ \\ \delta\tilde{s}_- \end{pmatrix} = \tfrac{1}{2} g_e \mu_B \begin{pmatrix} \chi_{+-} & 0 \\ 0 & \chi_{-+} \end{pmatrix} \begin{pmatrix} \tilde{b}_+ \\ \tilde{b}_- \end{pmatrix} \tag{55}$$

The transverse spin susceptibilities write:

$$\chi_{+-} = \frac{4\Pi_{\downarrow\uparrow}}{1-4G_{+-}^{xc}\Pi_{\downarrow\uparrow}} \quad \chi_{-+} = \frac{4\Pi_{\uparrow\downarrow}}{1-4G_{+-}^{xc}\Pi_{\uparrow\downarrow}} \tag{56}$$

The transverse spin response involves the excitation of electrons from an occupied state $|\mathbf{k},\sigma\rangle$ to an empty state $|\mathbf{k}+\mathbf{q},\bar{\sigma}\rangle$, where $\bar{\sigma}$ is the reverse spin state of $\sigma$. Thus, the energies of the spin-flip single particle excitations depend on the change in the kinetic energy and the renormalized Zeeman energy $Z^*$. The zeros of the denominators in the transverse spin susceptibilities are the spin-waves. Inside a spin-wave, the electron spins have a coherent movement which is a combination of transverse components oscillating at the spin wave eigenfrequency and the longitudinal spin component flipping at a frequency proportional to the spin-flip wave amplitude. Spin-flip excitations induce a dynamical change in the spin polarization degree.

In the unpolarized limit $\zeta$=0, we have $\Pi_{\uparrow\downarrow} = \Pi_{\downarrow\uparrow} = \tfrac{1}{2}\Pi$ and $G_{zz}^{xc} = 2G_{+-}^{xc}$. Therefore, the longitudinal and transverse spin responses match the following equalities:.

$$\chi_{-+}(\zeta=0) = \chi_{+-}(\zeta=0) = 2\chi_{zz}(\zeta=0) \tag{57}$$

### 1.i Local field factors scheme

Similar expressions for the above dynamical susceptibilities have been obtained using the local field factor formalism[15,16,17,43]. Such a description is in essence close to the spin density formalism as it describes the Coulomb correction to the non-interacting Hamiltonian of a single electron with spin $\sigma$, by quantities proportional to the induced densities[44]. The factor of proportionality is $V_q$ corrected from the corresponding local field factors $G_\sigma^+, G_{L,\sigma}^-$ and $G_{T,\sigma}^-$ for charge, spin density and spin-flip density perturbations[16], respectively. Local field factors are potentially a more complete description of dynamical screening as they can have wavevector and frequency dependence characteristic of non-locality effects contrary to the Kohn-sham potentials which are constant in space and time in a translationnaly invariant system. The difficulty relies in their derivation which has been addressed extensively in the literature[45,46] and is actually not completely solved. Concerning the spin-polarized 2DEG, Ref. 17 gives analytical expressions for spin-resolved local field factors in the small wave vector and static limit. These expressions were derived from the thermodynamic limits[46] followed by local field factors, which links them to the compressibility and spin-susceptibility of the equilibrium state, e. g. the one given in Eq. (12). The local field factors found in Ref. 17 are consequently equivalent to the potentials given in Eq. (50), except that the corrections due to transverse spin movement vanish in Eq. (12) and thus disappeared in the local field factors found in Ref. 17. In particular, a feature of the SP2DEG is the fact that the transverse potential $G_{+-}^{xc}$ differs from the longitudinal one $G_{zz}^{xc}$ (if $\zeta$=0, $G_{zz}^{xc} = 2G_{+-}^{xc}$), which means that $G_{L,\sigma}^-$ and $G_{T,\sigma}^-$ are different. This essential difference has not been considered in previous works of Refs. 15 and 17. A comparison of the response functions found in Ref. 17. with the ones given in Eqs (52) and (54), leads to the correct expressions for the thermodynamic limits of the local field factors:

$$
\begin{aligned}
G_{\sigma}^{+} &= -\frac{1}{V_{\mathbf{q}}}\left[G_{nn}^{xc} + \mathrm{sgn}(\sigma) G_{nz}^{xc}\right] \\
&= \frac{\tilde{q} r_s^2}{8\sqrt{2}F(q)}\left[\frac{\partial \varepsilon_{xc}}{\partial r_s} - r_s \frac{\partial^2 \varepsilon_{xc}}{\partial r_s^2} + 2\sigma \frac{\partial^2 \varepsilon_{xc}}{\partial r_s \partial \zeta} - 4\zeta \frac{\partial^2 \varepsilon_{xc}^{*}}{\partial r_s \partial \zeta} - \sigma \frac{4}{r_s}\frac{\partial \varepsilon_{xc}^{*}}{\partial \zeta} + 4\frac{\zeta}{r_s}(\sigma - \zeta)\frac{\partial^2 \varepsilon_{xc}^{*}}{\partial \zeta^2}\right] \\
G_{L,\sigma}^{-} &= -\frac{1}{V_{\mathbf{q}}}\left[G_{zz}^{xc} + \mathrm{sgn}(\sigma) G_{nz}^{xc}\right] \\
&= \frac{\tilde{q} r_s}{2\sqrt{2}F(q)}\left[-\left(1 - \sigma\zeta^{*}\right)\frac{\partial^2 \varepsilon_{xc}}{\partial \zeta^2} - \sigma \frac{\partial \varepsilon_{xc}^{*}}{\partial \zeta} + \sigma \frac{r_s}{2}\frac{\partial^2 \varepsilon_{xc}}{\partial r_s \partial \zeta}\right] \\
G_{T,\sigma}^{-} &= -\frac{2}{V_{\mathbf{q}}} G_{+-}^{xc} \\
&= -\frac{\tilde{q} r_s}{2\sqrt{2}F(q)}\frac{1}{\zeta}\frac{\partial \varepsilon_{xc}^{*}}{\partial \zeta}
\end{aligned}
\tag{58}
$$

where $\tilde{q} = q/k_F$ and additional terms not found in Ref. 17 are pointed by (*).

# 4. SPIN EXCITATIONS : DISPERSIONS AND SPECTRUM

## 1.j General considerations on the dissipation spectrum

Assuming that the SP2DEG is perturbed by only one of the two rotating fields $\mathbf{b}_{\pm}(\mathbf{r},t) = \left(\tilde{b}_x \mathbf{x} \pm i\tilde{b}_y \mathbf{y}\right)e^{i\mathbf{q}\cdot\mathbf{r} - i\omega t}$, the dissipation rate of the transverse response is given by[40]:

$$
W = -2\omega\left[\mathrm{Im}\,\chi_{\pm}(\mathbf{q},\omega)\right]\left|g_e \mu_B \tilde{b}_{\pm}\right|^2 \tag{59}
$$

For the longitudinal response, as we deal with a matrix response, the power dissipated will depend on the perturbing field imposed to the SP2DEG. For example, if only the oscillating potential $\varphi(\mathbf{r},t) = \tilde{\varphi}(q,\omega)e^{i\mathbf{q}\cdot\mathbf{r} - i\omega t} + c.c.$ is applied, the rate of energy-loss will be straightforwardly given by:

$$
W = -2\omega\left[\mathrm{Im}\,\chi_{nn}(\mathbf{q},\omega) + \mathrm{Im}\,\chi_{zn}(\mathbf{q},\omega)\right]\left|e\tilde{\varphi}\right|^2 \tag{60}
$$

where both charge and spin excitations contribute to dissipation. Equivalently, we would find the energy dissipated in the case of a unique magnetic field applied along the z direction, $\mathbf{b}(\mathbf{r},t)=\tilde{b}_z(q,\omega)e^{i\mathbf{q}\cdot\mathbf{r}-i\omega t}\mathbf{z}$, by permuting $n$ and $s_z$ indexes in Eq. (60), and replacing $|e\tilde{\varphi}|^2$ by $|g_e\mu_B\tilde{b}_z|^2$. As the typical range of energy considered is within a few meV, probing the dissipation through direct application of the longitudinal (transverse) fields can be performed by FIR spectroscopy (Electron paramagnetic resonance spectroscopy, EPR), where an infra-red photon linearly (circularly) polarized would be absorbed by longitudinal (transverse) excitations. Because of the momentum conservation law, such an absorption process is sensitive to vanishing-$q$ excitations only. Consequently, the longitudinal probe would be useless since intra-quantum well subband longitudinal excitations disappear at $q$=0. Standard EPR has also been unsuccessful in probing $q$=0 transverse spin excitations of SP2DEG because the typical amount of available spin per unit volume is close to the minimum experimental sensitivity even for which strong heating of the electrons through the microwave field does occur and prevents to resolve the response. Nevertheless, Raman spectroscopy has proven to be very powerful for probing zero and non-zero $q$ longitudinal and transverse excitations[9], as the two-optical- photon process allows a transfer of momentum $q$ to the SP2DEG. For typical densities ($r_s$=2) Raman transferred wave vectors $q$ ranging from 0 to $0.1k_F$ can be achieved in CdMnTe. Second order perturbation theory yields the Raman scattering rate which is standardly written as[47]:

$$\frac{d^2\sigma}{d\omega d\Omega}\propto\sum_M\left|\langle M|\sum_{\mathbf{k},\sigma,\sigma'}\gamma_{\sigma'\sigma}(\mathbf{q},\mathbf{k},\omega)c^{+}_{\mathbf{k}+\mathbf{q}\sigma'}c_{\mathbf{k}\sigma}|0\rangle\right|^2\delta(E_M-E_0-\hbar\omega) \qquad (61)$$

In Eq. (61), $c_{\mathbf{k}\sigma}$ ($c^{+}_{\mathbf{k}\sigma}$) are the destruction (creation) operators of an electron on state $|\mathbf{k},\sigma\rangle$; $|M\rangle$ and $|0\rangle$ are many body excited and ground states of the SP2DEG, with respective energy $E_M$ and $E_0$; $\gamma_{\sigma'\sigma}(\mathbf{q},\mathbf{k},\omega)$ is a coefficient involving a product of two optical matrix elements (one for each photon field) and a resonant denominator. In Zinc-blende host semiconductors like CdMnTe, the Raman cross section of Eq. (61) exhibits the following selection rules[47]: spin-conserving (resp. spin-flip) excitations are probed when the incoming and scattered photons have parallel (resp. crossed) polarizations. Thus, it is possible to separate the longitudinal and the transverse response. Moreover, to deal with Eq. (61), one commonly makes the rude assumption of neglecting the $\omega$, $\mathbf{k}$, $\mathbf{q}$ dependence (which implies suppressing the resonance) of the $\gamma$ factors. Modifications of the response introduced by resonant denominators have been discussed in Ref. 48. Therefore, Eq. (61) becomes:

$$\left(\frac{d^2\sigma}{d\omega d\Omega}\right)_{//} \propto \sum_M \left|\langle M|\left(\gamma_{\uparrow\uparrow}+\gamma_{\downarrow\downarrow}\right)\hat{n}_{\mathbf{q}}+\left(\gamma_{\uparrow\uparrow}-\gamma_{\downarrow\downarrow}\right)\hat{s}_{z\mathbf{q}}|0\rangle\right|^2 \delta\left(E_M-E_0-\hbar\omega\right) \quad (62)$$

for the polarized (parallel polarizations) case, and:

$$\left(\frac{d^2\sigma}{d\omega d\Omega}\right)_{\perp} \propto \left|\gamma_{\uparrow\downarrow}\right|^2 \sum_M \left|\langle M|\hat{s}_{+\mathbf{q}}|0\rangle\right|^2 \delta\left(E_M-E_0-\hbar\omega\right) \quad (63)$$

for the depolarized (crossed polarizations) geometry in case $\zeta<0$. To obtain Eqs. (62) and (63), we have made used of the second quantization expression of the Fourier transform of density operators $\hat{n}(\mathbf{r})$ introduced in Eq. (30) and spin operators $\hat{\mathbf{s}}(\mathbf{r})$ defined in Eq. (2):

$$\hat{n}_{\mathbf{q}}=\sum_{\mathbf{k},\sigma} c^+_{\mathbf{k}-\mathbf{q}\sigma}c_{\mathbf{k}\sigma},\ \hat{s}_{z\mathbf{q}}=\sum_{\mathbf{k},\sigma}\mathrm{sgn}(\sigma)\,c^+_{\mathbf{k}-\mathbf{q}\sigma}c_{\mathbf{k}\sigma},\ \hat{s}_{+\mathbf{q}}=\sum_{\mathbf{k}} c^+_{\mathbf{k}-\mathbf{q}\uparrow}c_{\mathbf{k}\downarrow} \quad (64)$$

By taking the imaginary part of the Lehman representation of the response functions and assuming a linear dependence on $\zeta$ of the quantity $(\gamma_{\uparrow\uparrow}-\gamma_{\downarrow\downarrow})/(\gamma_{\uparrow\uparrow}+\gamma_{\downarrow\downarrow}) = f(B_0) = g(\zeta) \approx \beta_2\zeta$, as we would naturally expect for spin dependant optical matrix elements, one finally finds:

$$\left(\frac{d^2\sigma}{d\omega d\Omega}\right)_{//} \propto \mathrm{Im}\,\chi_{nn}(\mathbf{q},\omega) + 2\beta_2\zeta\,\mathrm{Im}\,\chi_{nz}(\mathbf{q},\omega) + \beta_2^2\zeta^2\,\mathrm{Im}\,\chi_{zz}(\mathbf{q},\omega) \qquad (65)$$

and:

$$\left(\frac{d^2\sigma}{d\omega d\Omega}\right)_{\perp} \propto \mathrm{Im}\,\chi_{+-}(\mathbf{q},\omega) \qquad (66)$$

Thus, the Raman probe can separate the longitudinal and the transverse responses. For the unpolarized 2DEG, the parallel Raman response is given by $\mathrm{Im}\,\chi_{nn}$ only. In the following, we will consider the dispersions and dissipation spectrum of both longitudinal and transverse excitations.

### 1.k Single particle excitations

As said above, excitations appear as poles in the various linear response functions described in Eqs. (52) and (54). Some of these poles originate from poles of the Lindhart polarizabilities $\Pi_{\sigma\sigma'}$ themselves. They correspond to frequencies that cancel out the denominators in Eq. (46) where a single electron is excited from an occupied state $|\mathbf{k},\sigma\rangle$ to an empty state $|\mathbf{k}+\mathbf{q},\sigma'\rangle$. As there are many occupied states, these excitations form the single particle continuum (SPE), characterized by:

$$\hbar\omega_{SPE}^{\sigma\sigma'}(\mathbf{k},\mathbf{q}) = \varepsilon^*_{\mathbf{k}+\mathbf{q}\sigma'} - \varepsilon^*_{\mathbf{k}\sigma};\ \text{with } n_{\mathbf{k}\sigma} > 0 \text{ and } n_{\mathbf{k}+\mathbf{q}\sigma'} < 1 \qquad (67)$$

The occupancy conditions $n_{\mathbf{k}+\mathbf{q},\sigma'} < 1$ and $n_{\mathbf{k},\sigma} > 0$ define the boundaries of the continuum.

The spin-conserving continuum is a superposition of $\uparrow\rightarrow\uparrow$ and $\downarrow\rightarrow\downarrow$ SPEs. At zero temperature the boundary conditions for these SPE's are (see Fig. 3a):

$$\hbar\omega_{SPE}^{\sigma\sigma} = \hbar v_{F,\sigma} q + \hbar^2 q^2 / 2m_b\, ;\ \sigma = \uparrow, \downarrow \tag{68}$$

where we have introduced the 0K spin resolved Fermi velocities and wave vectors:

$$v_{F,\sigma} = \hbar k_{F,\sigma} / m_b \quad k_{F,\sigma} = k_F \sqrt{1 + \mathrm{sgn}(\sigma)\zeta} \tag{69}$$

The spin-flip continuum is also a superposition of $\downarrow\rightarrow\uparrow$ and $\uparrow\rightarrow\downarrow$ SPEs. In this case, the zero temperature boundary conditions are (see Fig. 3b):

$$\left.\begin{aligned} \hbar\omega_{SPE}^{\sigma\bar{\sigma}} &= Z^* \pm \hbar v_{F,\sigma} q + \hbar^2 q^2 / 2m_b \\ \hbar\omega_{SPE}^{\bar{\sigma}\sigma} &= -Z^* + \hbar v_{F,\bar{\sigma}} q + \hbar^2 q^2 / 2m_b \end{aligned}\right\} \sigma = \mathrm{sgn}\,\zeta, \bar{\sigma} = -\sigma \tag{70}$$

Spin-flip excitations with initial state in the minority spin population can exist only if $q$ is greater than a minimum wave vector:

$$q_0 = \left| k_{F,\downarrow} - k_{F,\uparrow} \right| \tag{71}$$

We note that for $q_0 < q < k_{F,\downarrow} + k_{F,\uparrow}$, it is always possible to have a zero energy spin-flip excitation, by keeping the electron on the Fermi disks.

### *Dissipation spectrum of single particle excitations*

In a preliminary step, we evaluate the energy dissipated through single particle excitations in an imaginary situation where all the dynamical screenings are switch off. This corresponds to setting equal to zero all the Kohn-Sham potentials of Eqs.(50) in the response functions. In such a situation, the SP2DEG response would reduce to a combination of Lindhartd-type polarizabilities. We find:

$$\begin{pmatrix} \delta\tilde{n} \\ \delta\tilde{s}_z \end{pmatrix} = \begin{pmatrix} \Pi_{\uparrow\uparrow} + \Pi_{\downarrow\downarrow} & \Pi_{\uparrow\uparrow} - \Pi_{\downarrow\downarrow} \\ \Pi_{\uparrow\uparrow} - \Pi_{\downarrow\downarrow} & \Pi_{\uparrow\uparrow} + \Pi_{\downarrow\downarrow} \end{pmatrix} \begin{pmatrix} -e\tilde{\varphi} \\ g_e \mu_B \tilde{b}_z \end{pmatrix} \text{ and } \begin{pmatrix} \delta\tilde{s}_+ \\ \delta\tilde{s}_- \end{pmatrix} = 4 \begin{pmatrix} \Pi_{\downarrow\uparrow} & 0 \\ 0 & \Pi_{\uparrow\downarrow} \end{pmatrix} \begin{pmatrix} \tilde{b}_+ \\ \tilde{b}_- \end{pmatrix} \tag{72}$$

Fig. 4 shows the Raman dissipation spectrum through single particle excitations only. For the spin-conserving case, using Eqs. (65) and (72), we find:

$$\left(\frac{d^2\sigma}{d\omega d\Omega}\right)_{//} \propto \left(1-\beta_2\zeta\right)^2 \operatorname{Im}\Pi_{\downarrow\downarrow}\left(q,\omega\right)+\left(1+\beta_2\zeta\right)^2 \operatorname{Im}\Pi_{\uparrow\uparrow}\left(q,\omega\right) \qquad (73)$$

And for the spin-flip, when $\zeta<0$:

$$\left(\frac{d^2\sigma}{d\omega d\Omega}\right)_{\perp} \propto \operatorname{Im}\Pi_{\downarrow\uparrow}\left(q,\omega\right) \qquad (74)$$

The former is the addition of two asymmetric lines, each of them is characteristic of the two-dimensional single particle spectrum associated to spin up and down populations. They respectively have a high energy cutoff at: $\hbar v_{F,\sigma} q + \hbar^2 q^2/2m_b$; $\sigma = \uparrow,\downarrow$. The minority population lines appear as a shoulder of the majority one (Arrows 3 and 4 in

Fig. 4).

The dip in the spin-flip spectrum is due to a reduction of the number of available excitations because of the phase space filling in the minority spin-subband. It disappears in the full polarized state.

### 1.l Collective excitations

Collective excitations appear as zeros in the denominators involved in the response functions. Hence, the present description supports the Fermi-liquid, where collective and single particle excitations do coexist. In the Fermi liquid description, SPEs are a memory of the non-interacting system, but they acquire a renormalized mass and a Zeeman splitting due to the short range term in the Coulomb interaction. The Coulomb interaction couples SPE having different initial state to build collective excitation. Consequently, SPEs have a finite lifetime while collective excitations can be damped by SPEs through the kinetic part of the SP2DEG Hamiltonian. In the adiabatic spin density approximation (ASDA) developed here at T=0K collective excitations are totally damped when they exactly enter the SPE continuum and are long-lived modes outside this continuum. Damping of spin waves occurring beyond the ASDA assumption has been recently considered in Ref. [49].

***Spin-conserving excitations: plasmons***

When finding the zeros of Eq. (53), only one is found outside the SPE continuum. An expansion of Lindhard polarizabilities around $q = 0$ leads to the analytic expression of its dispersion relation:

$$\tilde{\omega} = \left\{ \tilde{q} r_s 2\sqrt{2} + \tilde{q} r_s \sqrt{2} \left[ 2\sqrt{2} \tilde{k}_{F,\uparrow} \tilde{k}_{F,\downarrow} \left( G^-_{L,\downarrow} + G^-_{L,\uparrow} - G^+_{\uparrow} G^-_{L,\downarrow} - G^-_{L,\uparrow} G^+_{\downarrow} \right) - \sum_{\sigma} \tilde{k}_{F,\sigma} \left( G^+_{\sigma} + G^-_{L,\sigma} \right) \right] \right\}^{\frac{1}{2}} \qquad (75)$$

with $\tilde{\omega} = \hbar\omega / E_F$ , $\tilde{q} = q/k_F$ , $\tilde{k}_{F,\sigma} = k_{F,\sigma}/k_F$ and the local field factors of Eq. (58). The first term is exactly the well-known expression of the plasmon dispersion found in RPA (all G-factors set to zero). Hence, this longitudinal mode is a plasmon-like collective mode. Let us consider Fig. 5 where the dispersion of this mode has been numerically calculated for an ideal zero thickness quantum well. In Fig. 5a, one sees that exchange and correlations corrections introduce no qualitative change to the plasmon-like dispersion if compared to the RPA's one, only small quantitative changes are found for $\tilde{q}$ above 0.3. The exchange brings the strongest modification, specifically for high spin polarization degree, where exchange correction has a divergence. Correlations diminish the exchange effects. Fig. 5b shows the increasing depolarization shift (in units of $E_F$) with increasing $r_s$. The depolarization shift is the energy distance between the plasmon and the SPE boundary. This reveals a raise of the collective mode rigidity when reducing the density (Coulomb dominates over the kinetic energy). In Fig. 5c, we can see the total lack of sensitivity to the spin polarization degree of this plasmon-like mode, indicating that the spin-density component of the mode is negligible compared to the charge one.

***Charge density dissipation spectrum***

The dissipation spectrum through charge density given by $\mathrm{Im}\,\chi_{nn}$ has been plotted in Fig. 6(a) as a function of the spin-polarization degree $\zeta$ for a typical density $r_s$=2, width $w$=150Å, and fixed wave vector $\tilde{q} = q/k_F = 0.1$ accessible through Raman spectroscopy. The behavior of the SP2DEG $\mathrm{Im}\,\chi_{nn}$ has no additional features compared to the unpolarized 2DEG. It exhibits a plasmon peak rather insensitive to the spin-polarization degree. The presence of the dynamical screening ($V_\mathbf{q}$ and G-factors in Eqs.(52) and (53)) makes the plasmon show up while it captures the oscillator strength from the SPE, as illustrated by the inset of Fig. 6(a). Spin-conserving SPEs are almost totally screened out by the collective mode and have a negligible weight in the dissipation spectrum.

***Spin density dissipation spectrum***

Fig. 6(b) shows the dissipation spectrum through the spin density given by $\operatorname{Im}\chi_{zz}$ calculated for the same parameters. When the SP2DEG reaches the full polarized state ($\zeta$=-1), both the charge-charge and spin-spin responses have to coincide since only one spin population do exist. A spin-density contribution present in $\operatorname{Im}\chi_{zz}$ at intermediate $\zeta$, has to disappear at $|\zeta| = 1$ to let the plasmon mode capture all the oscillator strength as in $\operatorname{Im}\chi_{nn}$. On the other side, at $\zeta$ =0, $\operatorname{Im}\chi_{nn}$ and $\operatorname{Im}\chi_{zz}$ are decoupled, so that the plasmon mode do not appear in $\operatorname{Im}\chi_{zz}$. Consequently, the low energy features exhibited by $\operatorname{Im}\chi_{zz}$ in Fig. 6(b) are clearly attributed to spin-density-fluctuations. These spin excitations do not correspond to a long-lived spin-density wave mode as this is the case for the plasmon. Indeed, (1) at the limit $\zeta$=0, $\operatorname{Im}\chi_{zz}$ has no pole and (2) for finite $\zeta$, we find poles in the denominator of Eq. (53) lying in this energy range, but they do exist in the single particle domain and cannot exist as true collective modes. Nevertheless, the presence of these poles in the single particule domain gives a collective character to the SPE's. This become obvious when comparing the spin-fluctuations line shape (from $\operatorname{Im}\chi_{zz}$) and the SPEs line shape (see inset of Fig. 6(b)). Moreover, Fig.7(a) demonstrates that the lineshape deformation and red-shifting is more and more pronounced when increasing the coulomb interaction strength. Indeed, we compare the low energy part of $\mathrm{Im}\,\chi_{zz}$ with $\mathrm{Im}\,\Pi$ for various $r_s$ as a function of the reduced quantity $\tilde{\omega}$. In this picture, $\mathrm{Im}\,\Pi$ is a constant while the spin fluctuations line becomes narrower and experiences a red shift of its peak position whith incresing $r_s$. Dynamical screenings arising from both charge ($V_{\mathbf{q}}$, $G_{\sigma}^{+}$) and spin density ($G_{L,\sigma}^{-}$) are responsible for this behavior. In Fig. 7(b), (c) and (d), we seperate the role of charge and spin screenings. The spectrum is calculated for a typical value of $r_s$ ($r_s$=2) and various values of $\zeta$. In Fig. 7 (b), all dynamical screenings have been set to zero. $\mathrm{Im}\,\chi_{zz}$ is the addition of spin-conserving single particle excitations spectra associated to each spin population (see Eq.(72)). The minority spin SPE spectrum has a peak reaching zero at full polarization, while the majority one blue-shifts towards its final position that is equal to $\sqrt{2}$ times its position at $\zeta$=0. In Fig. 7(c), only charge dynamical screenings have been kept, this corresponds to keeping $V_{\mathbf{q}}$ and $G_{\sigma}^{+}$ but cancelling $G_{L,\sigma}^{-}$ in $\mathrm{Im}\,\chi_{zz}$. They introduce the plasmon mode in $\mathrm{Im}\,\chi_{zz}$ (shown in Fig. 6(b)) which captures the spectrum weight of the majority spin-conserving SPE (destroyed in Fig. 7(c)), while blue-shifting the minority ones. The line shape of the latter, except for the peak shifting, has a negligible deformation. The more striking behavior is obtained when activating the screening due to spin density (keeping $G_{L,\sigma}^{-}$ in $\mathrm{Im}\,\chi_{zz}$) as shown in Fig. 7(d). The spin-fluctuations spectrum becomes is red-shifted from the minority SPE spectrum and displays enhanced weights on the low energy side. When $|\zeta| \to 1$, the spin-fluctuations disappear from the spectrum as explained before. Hence, we demonstrate that collective effects determine the behavior of the spin-fluctuation spectrum. The shift evidenced in Fig. 7(a) has been observed in depolarized Raman spectra obtained from unpolarized 2DEG[3] and was already known in paramagnetic metals ($\zeta$~0) as the paramagnon effect[50]. $\mathrm{Im}\,\chi_{zz}$ can indeed be observed in depolarized Raman spectra, only for $\zeta$=0, where the spin degeneracy and isotropy is kept. In such case, $\chi_{zz}$ is equivalent to the transverse response $\chi_{+-}$. The behavior shown in Fig. 7(d) is a special feature of the SP2DEG and has been primarily observed in polarized Raman measurements carried out on 2DEG embedded in a $Cd_{0.992}Mn_{0.008}Te$ quantum well[10].

### *Longitudinal Raman spectrum*

Dissipation through intra-subband charge density excitations has been, in the past, extensively studied in unpolarized 2DEG. Raman scattering has been successfully employed on unpolarized 2DEG embedded in a doped single quantum well[3]. More recently it has been shown[51] that the intermediate state involved in the two-photon Raman process plays a role in the experimental polarized Raman spectra: when photons are in resonance with excitonic transitions, the experimental spectra support the theoretical expression of Eq. (65) calculated for $\zeta$=0 ($\propto \mathrm{Im}\,\chi_{nn}$), in which, the plasmon mode carries most of the spectral weight (see $\mathrm{Im}\,\chi_{nn}(\zeta=0)$ in Fig. 6). In strong resonance with the Fermi edge absorption however, the low energy SPE contributions show up in the Raman spectra. They carry a weight $10^3$-$10^4$ times stronger than the predicted weight in $\mathrm{Im}\,\chi_{nn}$. The origin of this discrepancy has been solved[48] by calculating the full resonant Raman response (keeping the resonant denominators in Eq. (61)). The strong resonance situation adds an additional coupling between light and SPEs which compete with the Coulomb coupling between SPEs. Dynamical screening of SPEs becomes uneffective and scattering through SPE is restored. Therefore, the Raman spectra reveal poles of both unscreened $\mathrm{Im}\,\Pi$ and $\mathrm{Im}\,\chi_{nn}$, with comparable weights until Coulomb dominates (high $r_s$). This works for the unpolarized 2DEG where the collective plasmon mode is out of the SPE domain. We might expect a different behavior for the SP2DEG where the low energy excitations (spin density fluctuations) are neither spin-conserving SPEs, nor a collective spin-density mode, but a mixture of both types as explained above. As the spin-conserving resonant Raman response is out of the scope of the present work, we show in Fig. 8 the longitudinal, non-resonant, Raman dissipation spectrum of Eq. (65). The polarized Raman response is in general a combination of the charge-charge, spin-spin and spin-charge dissipation spectrum, except for $\zeta$=0, where it couples only to the charge-charge response. As the spin-charge dissipation spectrum has spin-fluctuations and plasmon weights of opposite sign, it results in two situations which might depend on the sign and amplitude of the coefficient $\left(\gamma_{\uparrow\uparrow}-\gamma_{\downarrow\downarrow}\right)/\left(\gamma_{\uparrow\uparrow}+\gamma_{\downarrow\downarrow}\right)=f\left(B_0\right)$. This coefficient is necessary a function of the magnetic field and vanishes at zero field. For the sake of simplicity, it has been assumed to vary as $\beta_2\zeta$. We think this variation captures the qualitative behavior of the theoretical Raman spectra. The sign of $\beta_2$ depends on the hole state involved in the optical matrix elements in $\gamma_{\uparrow\uparrow}$ and $\gamma_{\downarrow\downarrow}$ and the resonant denominators. For positive (resp. negative) $\beta_2$, as shown in Fig. 8(a) (resp. Fig. 8(b)), the plasmon peak has a weight increasing (resp. decreasing) with the spin polarization degree, as a consequence of the constructive (resp. negative) interference between $\mathrm{Im}\,\chi_{nn}+\left(\beta_2\zeta\right)^2\mathrm{Im}\,\chi_{zz}$ and $\mathrm{Im}\,\chi_{nz}$. These behaviors have yet not been observed.

***Spin-flip excitations: spin-flip wave***

Spin-flip excitations appear as poles of the transverse spin susceptibilities $\chi_{-+}$ and $\chi_{+-}$ given in Eq. (56). Since $\chi_{-+}\left(-\omega,-\zeta\right)=\chi_{+-}\left(\omega,\zeta\right)$, we will restrict the discussion to positive poles of $\chi_{+-}\left(\omega,\zeta\right)$ found for $\zeta<0$. These are again formed by the poles of the unscreened susceptibility $\Pi_{\downarrow\uparrow}$, which are the spin-flip single particle excitations (SF-SPE) discussed in 1.k and the collective modes given by the zeros of the denominator in Eq. (56):

$$1-4G_{+-}^{xc}\Pi_{\downarrow\uparrow}\left(q,\omega\right)=0 \qquad (76)$$

Depending on $q$, Eq. (76) has one or none solution. The unique solution corresponds to the spin-flip wave (SFW) mode propagating below the SF-SPE continuum (see Fig. 9). A small-q expansion of $\Pi_{\downarrow\uparrow}$ leads to its zone center dispersion:

$$\hbar\omega_{SFW}\left(q\to 0\right)=Z-\frac{1}{|\zeta|}\frac{Z}{Z^{*}-Z}\frac{\hbar^{2}}{2m_{b}}q^{2}=Z\left[1-\frac{1}{r_{s}^{2}|\zeta|}\left(\frac{\partial\varepsilon_{xc}}{\partial\zeta}\right)^{-1}\tilde{q}^{2}\right] \qquad (77)$$

where we have made use of Eqs. (17) and (29). We remind that in Eq. (77) $Z^{*}$ is an approximated quantity containing the mass renormalization, while the ratio $Z^{*}/Z$ has to be replaced by the exact expression of Eq. (29). We thus prefer the right hand side of the equality (77) which does not depend on the approximation used for the self-energy. The SFW mode propagates parallel to the plane of the well. The Coulomb interaction between mobile electron spins is responsible for its propagation. If Coulomb interaction was switched off, Eq. (76) would have no solution. A key property of the SFW dispersion is that: $\hbar\omega_{SFW}\left(q=0\right)=Z$, an energy which does not depend on the Coulomb interaction between electrons. The ensemble spin motion ($q$=0) is defined by the external magnetic action only, a property also enounced as the *Larmor theorem*[52], a consequence of the rotational invariance of the spin degrees of freedom. The *Larmor's theorem* is the spin equivalent of the *Kohn theorem*[53] applying to the orbital degrees of freedom under translational invariance. This property enables a high accuracy determination of both Mn-electron temperature and Mn concentration $x$ by measuring the zone center energy of the spin flip wave[9]. We can conclude that the ASDA formalism employed in section 1.f supports this high symmetry theorem.

Fig. 9 reproduces the SFW dispersions obtained by solving Eq. (76) for typical $r_s$ values. The SFW mode propagates below the SF-SPE continuum in a wave vector window ranging from $q$=0 to $q$=$q_0$, point for which zero-energy SF-SPEs are present. When $|\zeta|$ increases from 0 to 1, the window enlarges and reaches $q=\sqrt{2}k_F$. There is no qualitative change of the dispersion between intermediate and full polarization. A surprising fact is the negative slope of the SFW dispersion which always lies below the low energy boundary of the SF-SPE continuum. This behavior, similar to the SFW in Landau quantized 2DEG, has led several authors to name the SFW a triplet exciton mode[54]. This name is meaningful for SF-SPEs pinned to discrete inter-Landau excitation energy. However, it becomes dubious in the SP2DEG case because it may suggest that the SFW is bound to the SF-SPE lowest energy branch, corresponding to the kinetic excitation of electrons occupying states in the vicinity of $-k_{F,\uparrow}$. A better understanding for the SFW energy shift $\Delta = \hbar\omega_{SFW}\left(q\right) - Z$ is the following: the Pauli hole, consequence of the Pauli repulsion between electrons, has a radius wider for parallel spins than for anti-parallel spins[55]. Hence, a reduction of the Coulomb energy between electrons occurs and is stronger for parallel spins than for anti-parallel spins. The Pauli hole is responsible for the enhancement and stability of the spin-polarization. Indeed, the larger the amount of aligned spins the lower the Coulomb energy. Consequently, the excitation of the polarized ground state by a coherent flip of all the spins (*q*=0 SFW mode) induces no change of the Pauli repulsion. Hence, no Coulomb corrections contribute to such a mode: $\hbar\omega_{SFW}(q=0)=Z$. On the contrary, flipping the spin of a single electron in the majority spin subband while keeping the others undisturbed, modifies the Pauli hole around the flipped electron. All the electrons of the majority spin-subband are redistributed to make the Pauli hole smaller leading to an increased Coulomb energy. Consequently, exciting a *q*=0 SF-SPE is more costly in energy: $Z^* \geq Z$. For $q \neq 0$, spins in the SFW mode are periodically anti-parallel for each $\lambda = \pi/q$. Compared to the *q*=0 situation, this induces a reduction of the Coulomb energy more and more pronounced when $\lambda$ is shorten, resulting in a negative slope dispersion.

***Spin-flip dissipation spectrum***

In Fig. 10 we have plotted the typical $\mathrm{Im}\,\chi_{+-}$ spectrum with parameters identical to the ones in Fig. 9(a), except for the addition of temperature $\tilde{T}=0.02$ and disorder $\tilde{\eta}=0.02$. The SFW peak clearly appears below the SF-SPE continuum and captures most of the oscillator strength from the SF-SPEs. The latter are strongly screened and have lost their specific line shape (see

Fig. 4(b)). As we are dealing here with finite temperature and disorder, the SFW mode starts to be damped by SF-SPEs before entering the 0K SF-SPE low energy boundary, the SFW peak acquires a high energy tail, a signature of the lost of its collective nature. Fig. 10(b) exhibits the $q$=0 peak evolution when increasing $\zeta$. The total weight carried by the SFW peak (as shown in the inset of Fig. 10(b)) varies linearly with the spin-polarization degree according to the exact sum rule:

$$\int_0^\infty \mathrm{Im}\,\chi_{+-}\left(q=0,\omega\right)d\omega = \frac{4m_b}{\hbar^2}\zeta \qquad (78)$$

## 5. CONCLUSION

We have shown how an electron gas embedded in magnetic quantum wells like $Cd_{1-x}Mn_xTe$/CdMgTe can generate a model SP2DEG, and we have presented conditions to obtain it by investigating the fundamental parameter characterizing the ground state : the spin-polarization degree. A domain emerges in the electron and Mn concentration plane where the full polarized state can be achieved. Furthermore, we have developed a formulation of the SP2DEG dynamics by calculating the response functions in the ASDA formalism. It required little adaptation of a formalism presented by A.K. Rajagopal in Ref. 20. The giant Zeeman energy of this magnetic system and the resulting high spin polarization degree decouples longitudinal and transverse spin motion. The long wavelength limit of the transverse local filed factor is derived from the above formalism. Calculations included a recent evaluation of the correlation energy[14] corrected from finite thickness of the well[36]. Having the response functions in hand, we have presented a full description of the charge and spin excitations, individual and collective, of the SP2DEG together with their dissipation spectrum. The particular case of the Raman spectrum has been discussed in detail. Two main novelties were shown: (1) the polarized Raman spectrum obtained for finite spin polarization degree reveals the spin-density fluctuations spectrum which behavior is dominated by collective effects, a point which was not discussed in the past. (2) We give emphasis on the fact that, due to the DMS giant Zeeman effect, the SP2DEG is an original model situation: a highly polarized two dimensional paramagnetic conducting system embedded in a semiconductor heterostructure whose physics resembles that of a paramagnetic metal except that the spin polarization degree is here comparable to that of a ferromagnetic metal. In the latter, magnetic excitations are zero sound spin waves[56] and Stoner excitations[57] centered at energy $2E_F\zeta$ in the eV range, which can be probed by spin polarized electron energy loss spectroscopy[Erreur ! Signet non défini.]. The SP2DEG has an energy that renders its transverse dynamics accessible to resonant Raman scattering already employed on two dimensional semiconductor structures. To summarize, novelties of the SP2DEG allows the full understanding of the spin-spin density and the transverse response functions.

The author would like to thank D. R. Richards, B. Jusserand and G. Vignale for fruitful discussions.

## Figures and captions

**Fig. 1** (Color online) Zeeman energy of Eq. (10) for $Cd_{0.99}Mn_{0.01}Te$ and various temperatures.

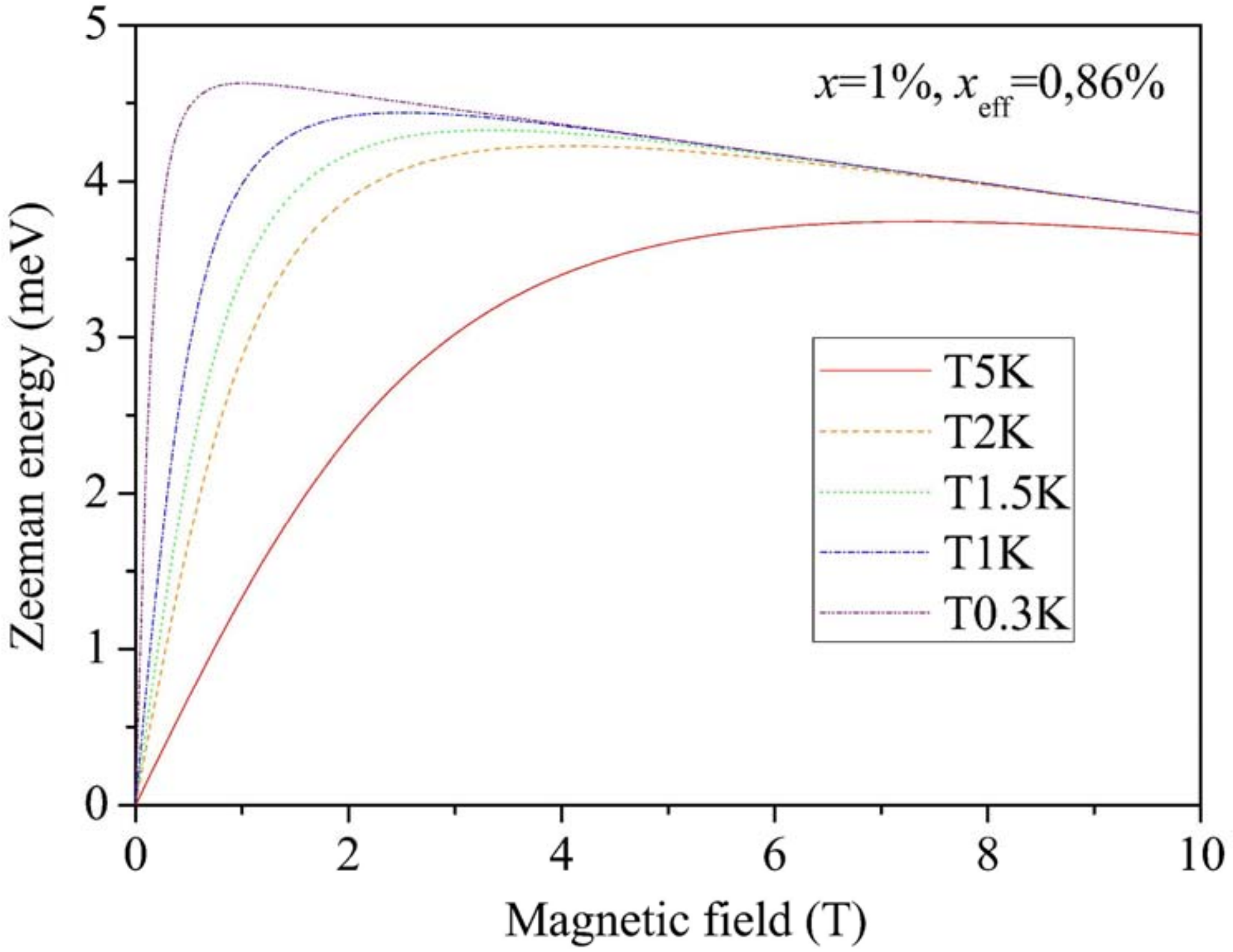

**Fig. 2** (Color online) Map of the maximum equilibrium spin polarization degree $\zeta = (n_\uparrow - n_\downarrow)/(n_\uparrow + n_\downarrow)$ as a function of the electron sheet density $n_{2D}$ and nominal Mn concentration $x$ for a typical $Cd_{1-x}Mn_xTe$/CdMgTe quantum well of width $w$=150Å. The Zeeman energy is taken at its maximum value (maximization of Eq. (10)) for temperature $T$=1.5K. (a) Bare spin polarization degree $\zeta_0$ of Eq. (15). (b) Calculation of $\zeta$ with interaction enhancement of Eq. (16)

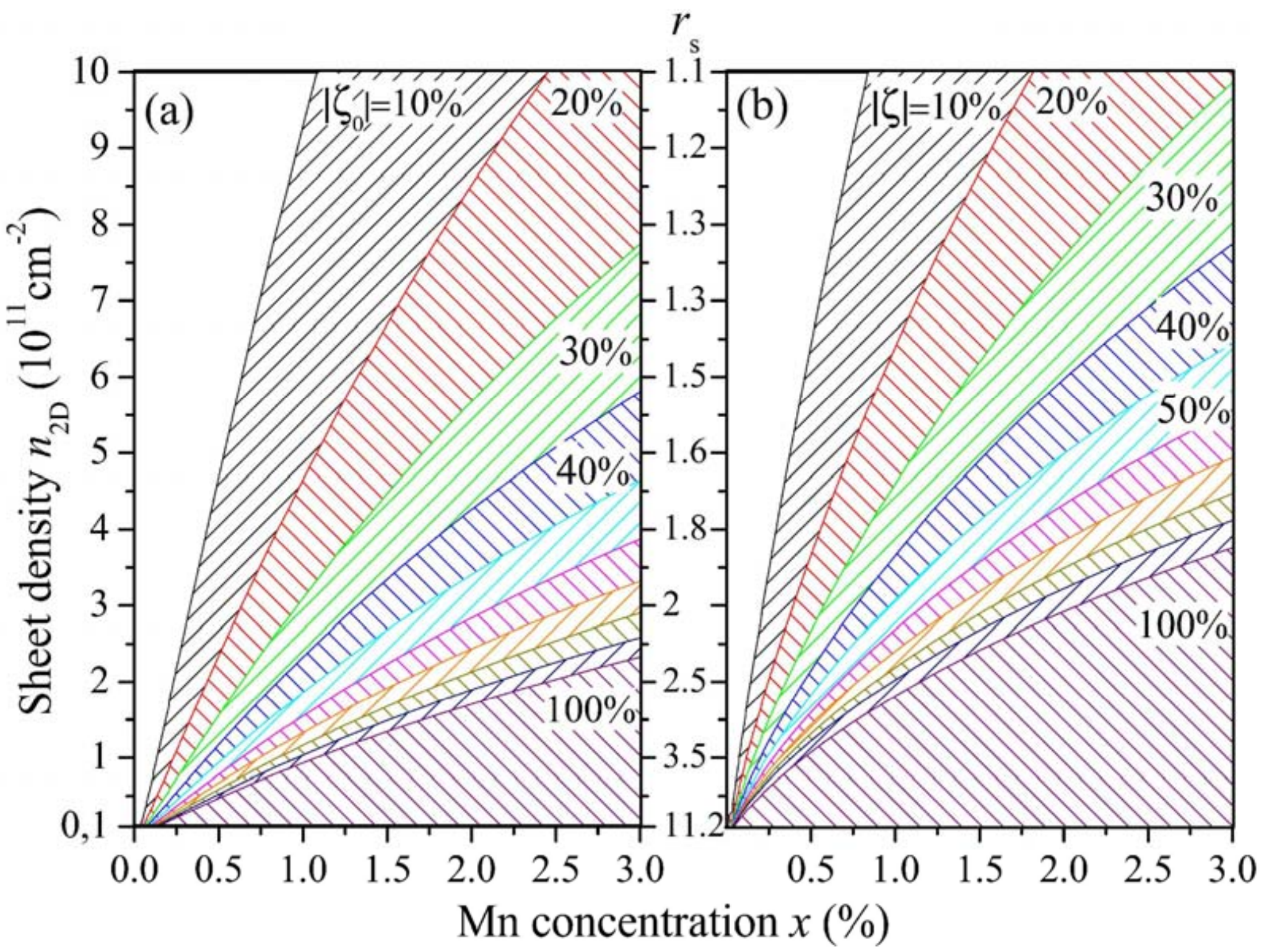

**Fig. 3** Map of the SPE continua for $\zeta$=-40% and $T$=0K. (a) The spin-conserving SPE continuum (SC-SPE) is the overlap of majority and minority SC-SPE continua. The high energy boundary of the former (the later) is indicated by line 4 (line 2). These two lines correspond to the particular SPEs indicated by arrows 4 and 2 in the inset. For SPEs having an initial state in between 1 and 2 (resp 3 and 4), no phase space reduction due to the final state occupancy does occur. SPEs with energies below line 1 (resp. line 3) have in contrary a reduced final state phase space. Hence, lines 1 and 3 indicate the peak position in the joint density of excitations (peak in $\mathrm{Im}\,\Pi_{\uparrow\uparrow}(q,\omega)$ and $\mathrm{Im}\,\Pi_{\downarrow\downarrow}(q,\omega)$ respectively). (b) Spin-flip SPE continuum (SF-SPE): lines 1 and 4 mark the boundaries of the $\downarrow\rightarrow\uparrow$ SPE continuum corresponding to excitations of arrows 1 and 4 in figure (c). The low energy boundary 1 reaches 0 for $q = q_0 = k_{F,\downarrow} - k_{F,\uparrow}$. For $q_0 \le q \le k_{F,\downarrow} + k_{F,\uparrow}$, it is always possible to find a zero energy SPE. Lines 2 and 3 indicate the peak position of the joint density of excitations (peaks in $\mathrm{Im}\,\Pi_{\downarrow\uparrow}(q,\omega)$) due to the spin up phase space that is reduced (arrows 2 and 3). Line 5 is the boundary of the $\uparrow\rightarrow\downarrow$ SPE continuum corresponding to excitation of arrow 5 in figure (d). Line 6 indicates the peak position of the excitation count occurring because of the spin down states occupancy (arrow 5). $\uparrow\rightarrow\downarrow$ SPE require a minimum wave vector $q_0$ to exist.

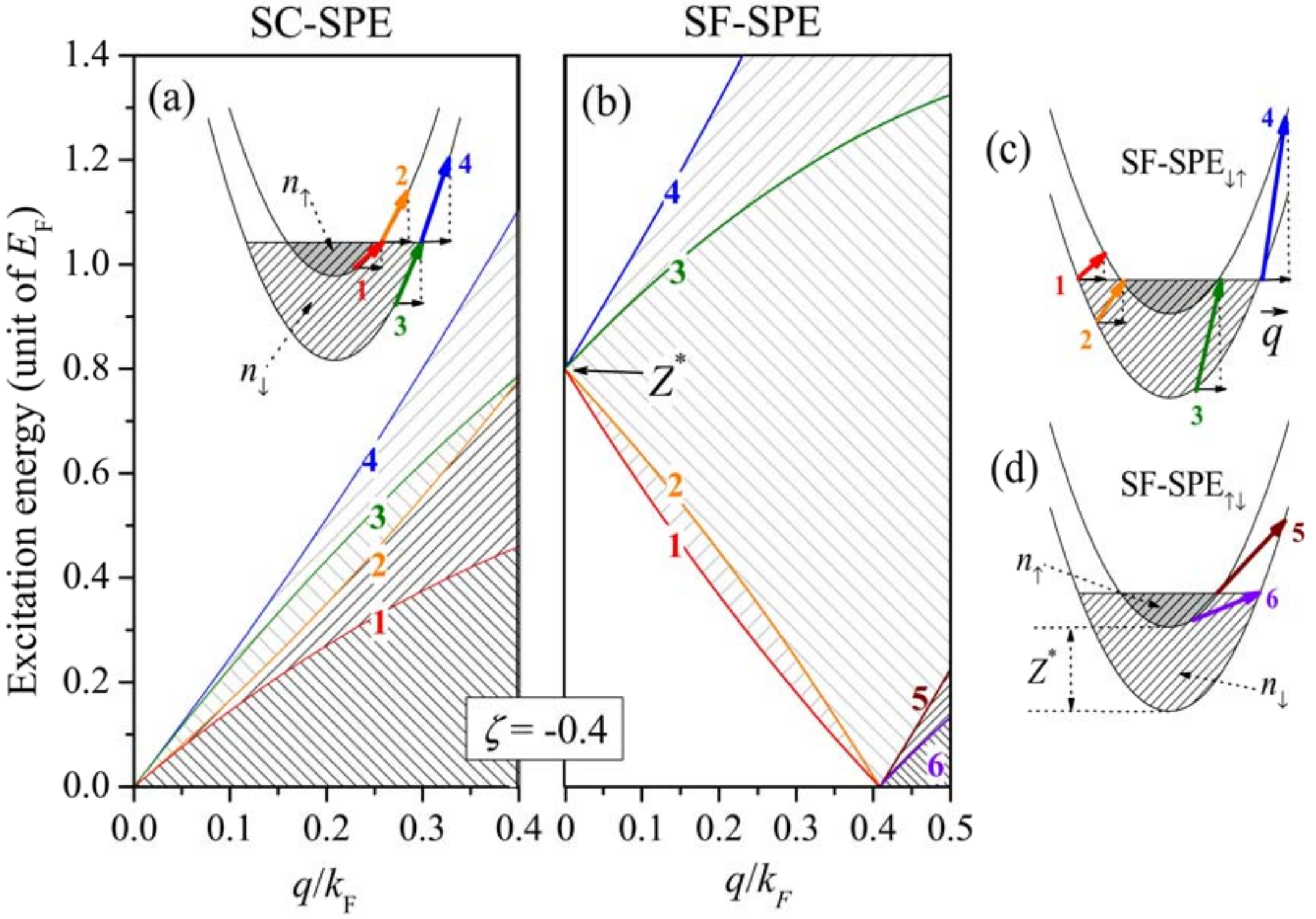

**Fig. 4** (Color online) Single particle Raman dissipation spectrum as a function of wave vector $q$ for a given spin polarization degree $\zeta$. Calculations have been performed for T=1.5K and homogeneous broadening $\hbar\eta = 0.02E_F$ . **(a)** Longitudinal spectrum of Eq.(73) with $\beta_2$=1. **(b)** Transverse spectrum of Eq. (74). Arrows 1,2,3,4 correspond to excitations pointed in Fig. 3.

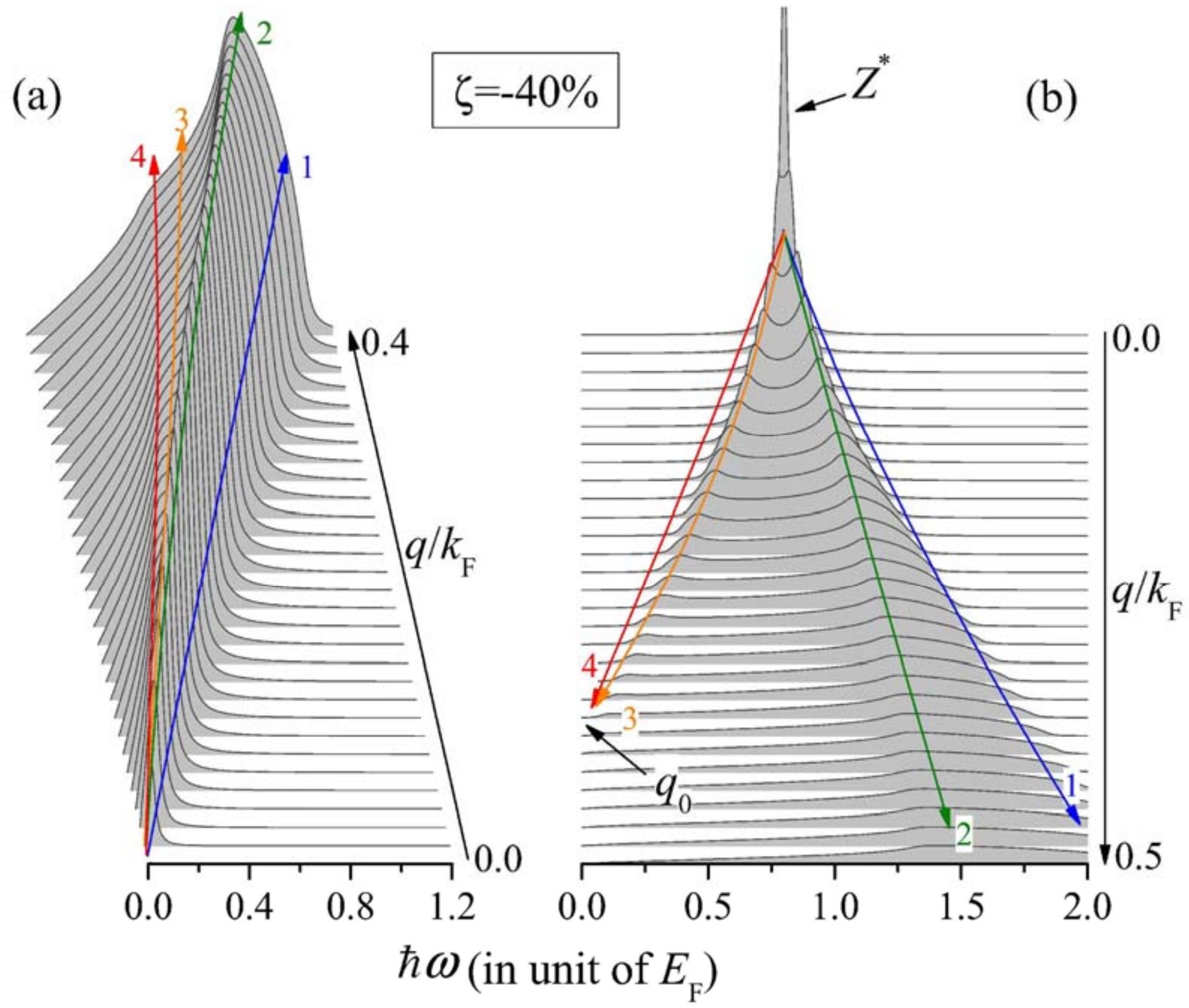

**Fig. 5** (Color online) Plasmon dispersions : **(a)** Effect of corrections : without exchange and correlations=RPA (straight line), with exchange only (dotted line), and with exchange and correlations (dashed line). **(b)** Effect of density : calculations are carried out with exchange and correlations for $r_s$=0.4, 1, 2, 5, 10. **(c)** Effect of spin polarization degree: variation of the plasmon dispersions for $\zeta$ ranging from 0 to -1, the shaded area are the intermediate values. In (a) and (b) the $SPE_{\uparrow\uparrow}$ and $SPE_{\downarrow\downarrow}$ boundaries are plotted for T=0K. (sparse domains). In (c) only the $SPE_{\downarrow\downarrow}$ boundary is plotted for $\zeta$=0 and $\zeta$=-1.

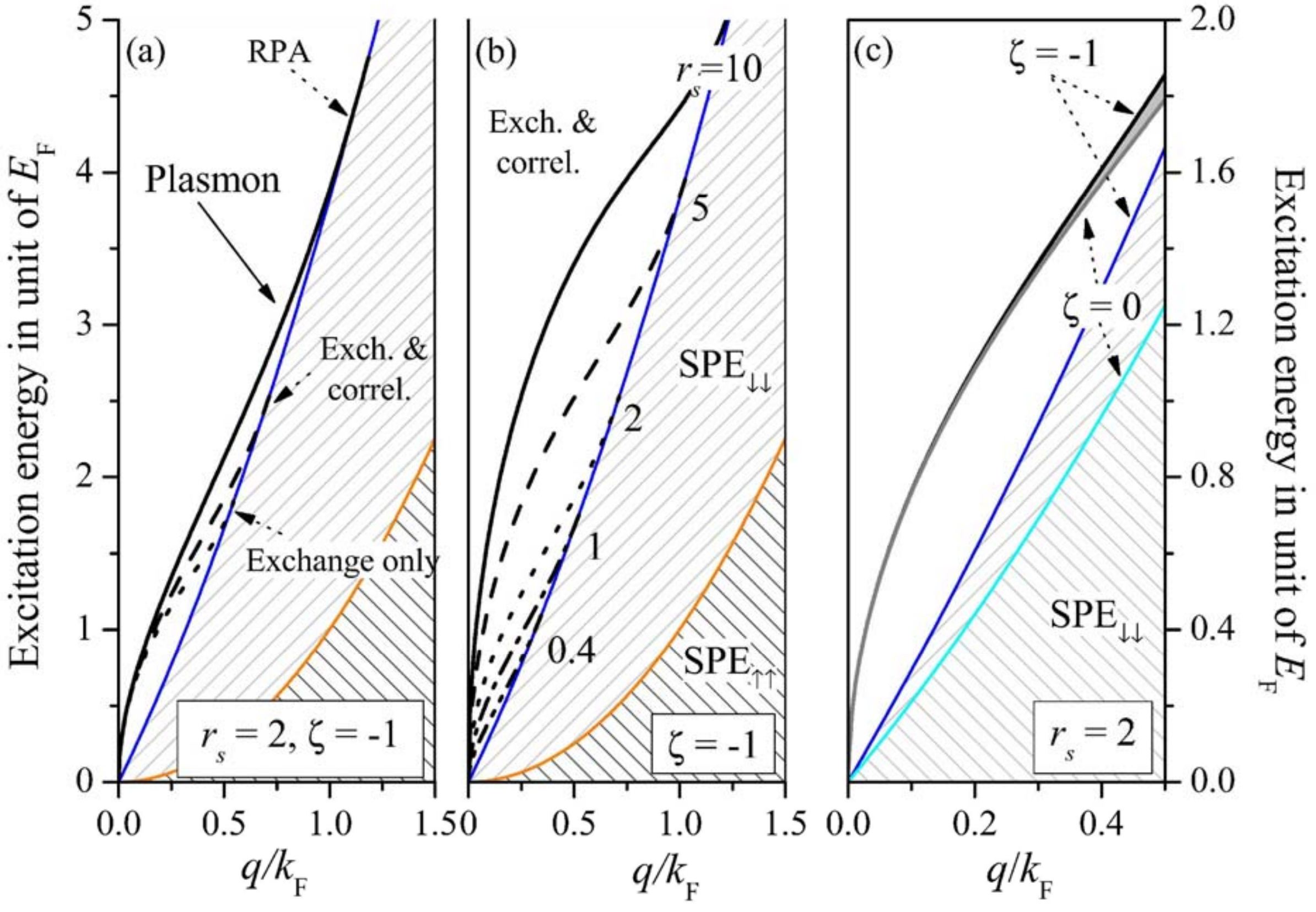

**Fig. 6** (Color online) Dissipation spectrum of spin-conserving response functions as a function of the spin polarization degree $\zeta$ and for $r_s$=2, wave vector $\tilde{q}=0.1$, homogenous broadening $\tilde{\eta}=\hbar\eta/E_F=0.02$, temperature $\tilde{T}=kT/E_F=0.02$ and $\tilde{\omega}=\hbar\omega/E_F$. (a) Charge-charge density response function with its plasmon peak; inset: illustration of the SPE screening, the region of SPE has been magnified. (b) Spin-spin density response function exhibiting a low energy spin-fluctuation part with a plasmon peak appearing at high $\zeta$; inset : illustration of the shape difference between the SPE dissipation and spin-fluctuations part: this part is a/the collective effect. (c) Charge-spin density response functions (extra-diagonal term in the longitudinal response). Both spin-fluctuations and plasmon contribute with growing weight as the spin-polarization degree increases.

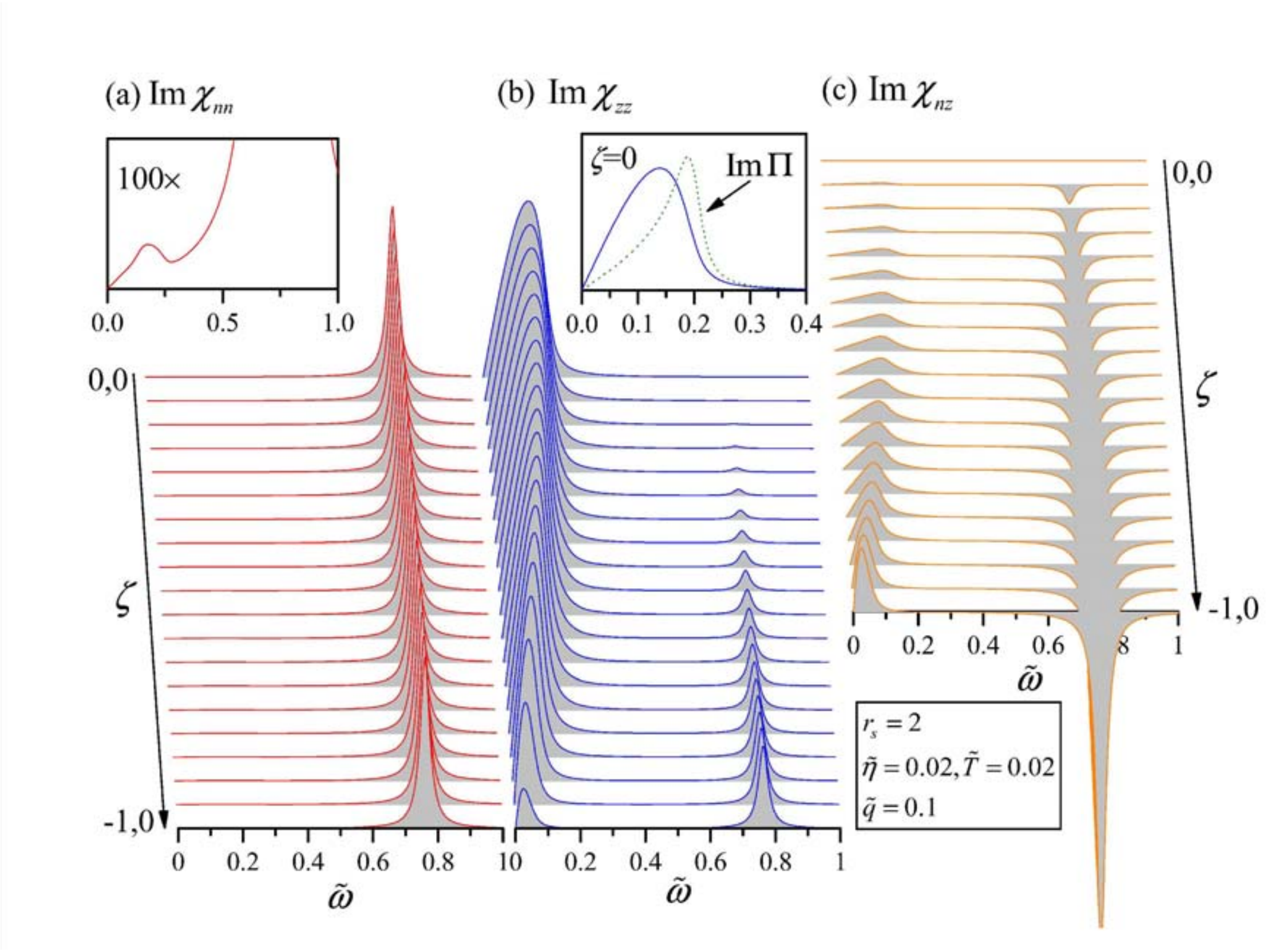

**Fig. 7** (Color online) Collective behavior of the spin fluctuations part in the spin-spin density response function. Calculations have been made at $\tilde{q} = 0.1$.**(a)** Lineshape deformation and peak shifting of the spin fluctuations as a function of $r_s$ for $\zeta$=0, $\tilde{\eta} = 0$ and $\tilde{T} = 0$. In this reduced unity scale, the SPE line in $\mathrm{Im}\,\Pi$ remains constant. **(b)**, **(c)** and **(d)** Origin of the spin-fluctuations lineshape deformation, $\mathrm{Im}\,\chi_{zz}$ is calculated at $r_s$=2 as a function of spin degree $\zeta$, by modifying the dynamical screening: (b) all the dynamical screenings are set to zero, (c) screening due to spin-density ($G^{-}_{L,\sigma}$) set to zero, (d) no modification.

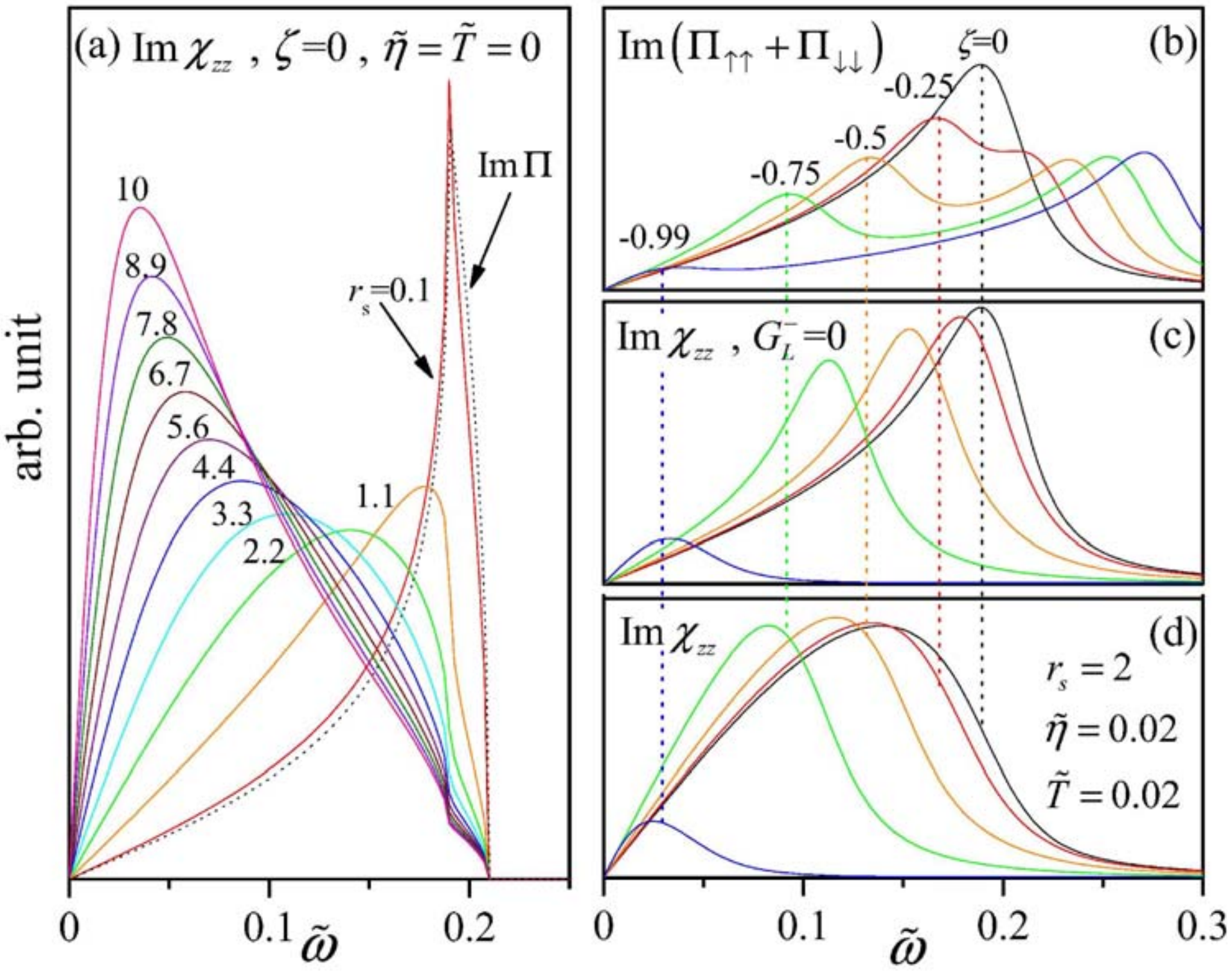

**Fig. 8** (Color online) Behavior of the longitudinal Raman dissipation spectrum. **(a)** Case with $(\gamma_{\uparrow\downarrow}-\gamma_{\downarrow\downarrow})/(\gamma_{\uparrow\downarrow}-\gamma_{\downarrow\downarrow})\le 0$. **(b)** Case with $(\gamma_{\uparrow\downarrow}-\gamma_{\downarrow\downarrow})/(\gamma_{\uparrow\downarrow}-\gamma_{\downarrow\downarrow})\ge 0$.

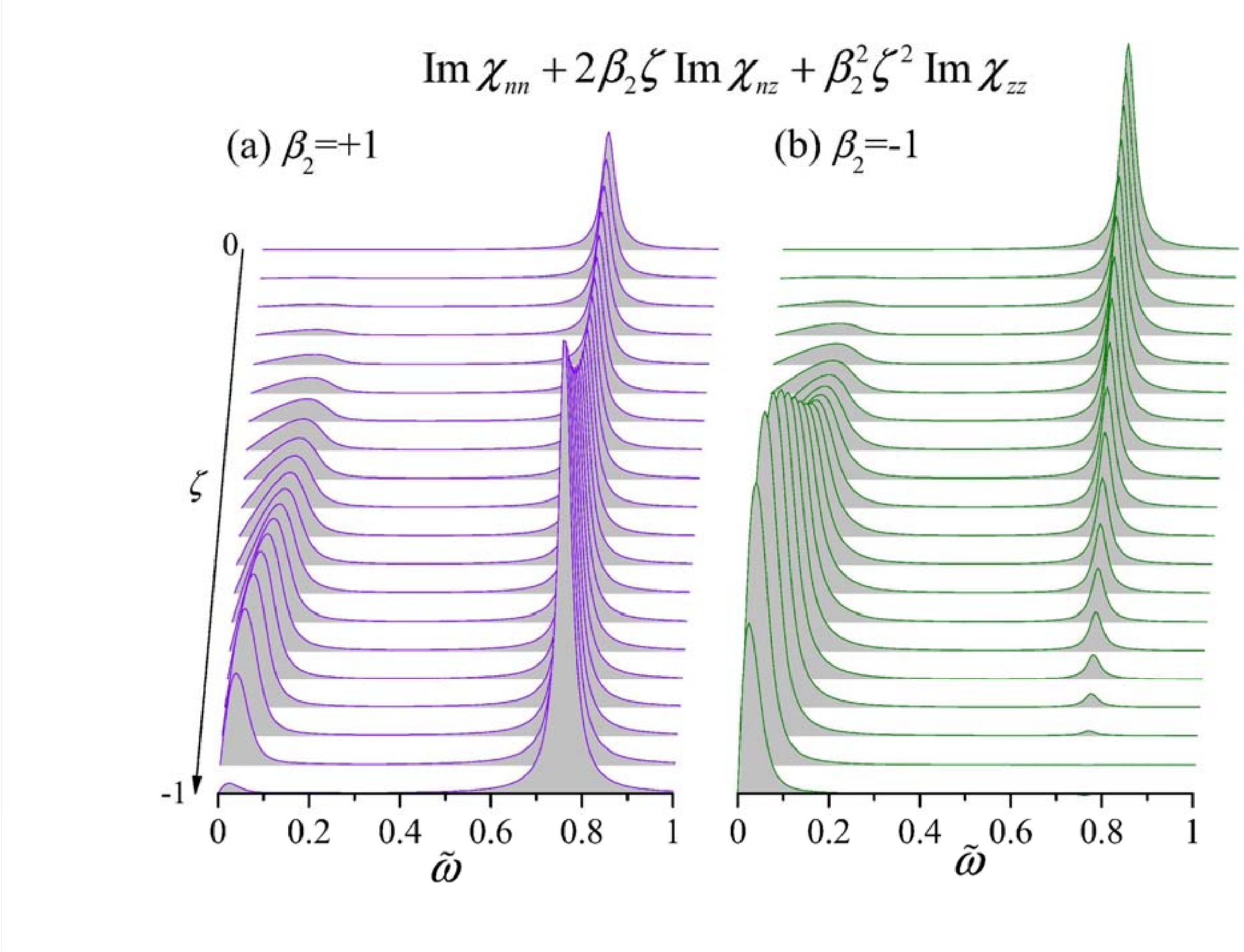

**Fig. 9** (Color online) Dispersion of spin flip excitations calculated for a zero thickness quantum well at T=0K, $\eta$=0 and $\zeta$=-0.4 **(a)** or $\zeta$=-1 **(b)**. Lines are the SFW mode corresponding to the indicated $r_s$ values. Domains are boundaries of the SF-SPE continua (which does not depend on $r_s$ in the reduced unity frame) and $\tilde{q}_0 = \sqrt{1-\zeta} - \sqrt{1+\zeta}$ .

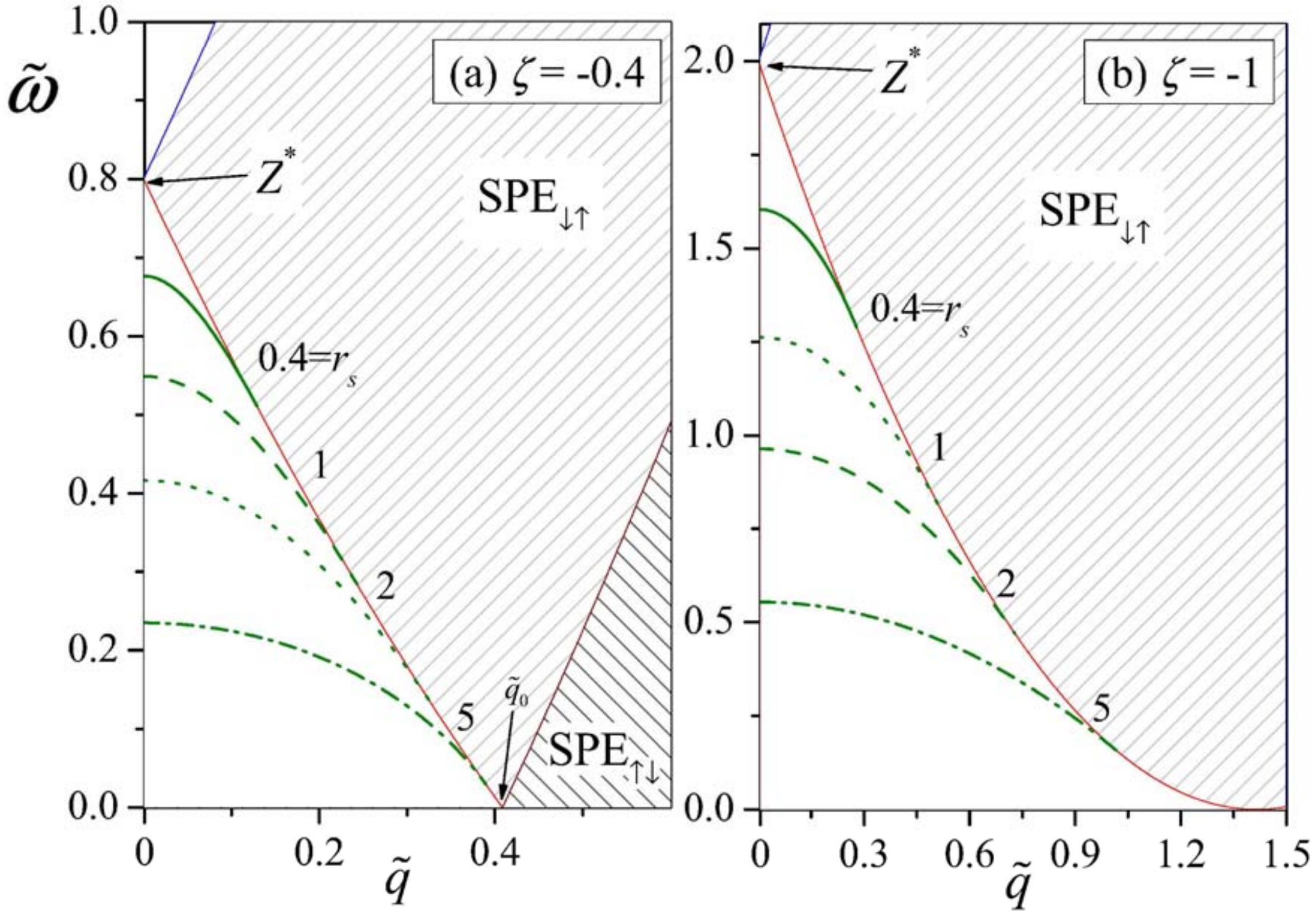

**Fig. 10** (Color online) Spin-flip dissipation spectrum $\mathrm{Im}\,\chi_{+-}$ calculated for $r_s = 2,\ \tilde{\eta} = 0.02,\ \tilde{T} = 0.02$ ; **(a)** as a function of the wave vector for *ζ*=-0.4. **(b)** as a function of *ζ*, for *q*=0. The inset of (b) plots the weight of the SFW peak deduced from the spectrum, as a function of *ζ*.

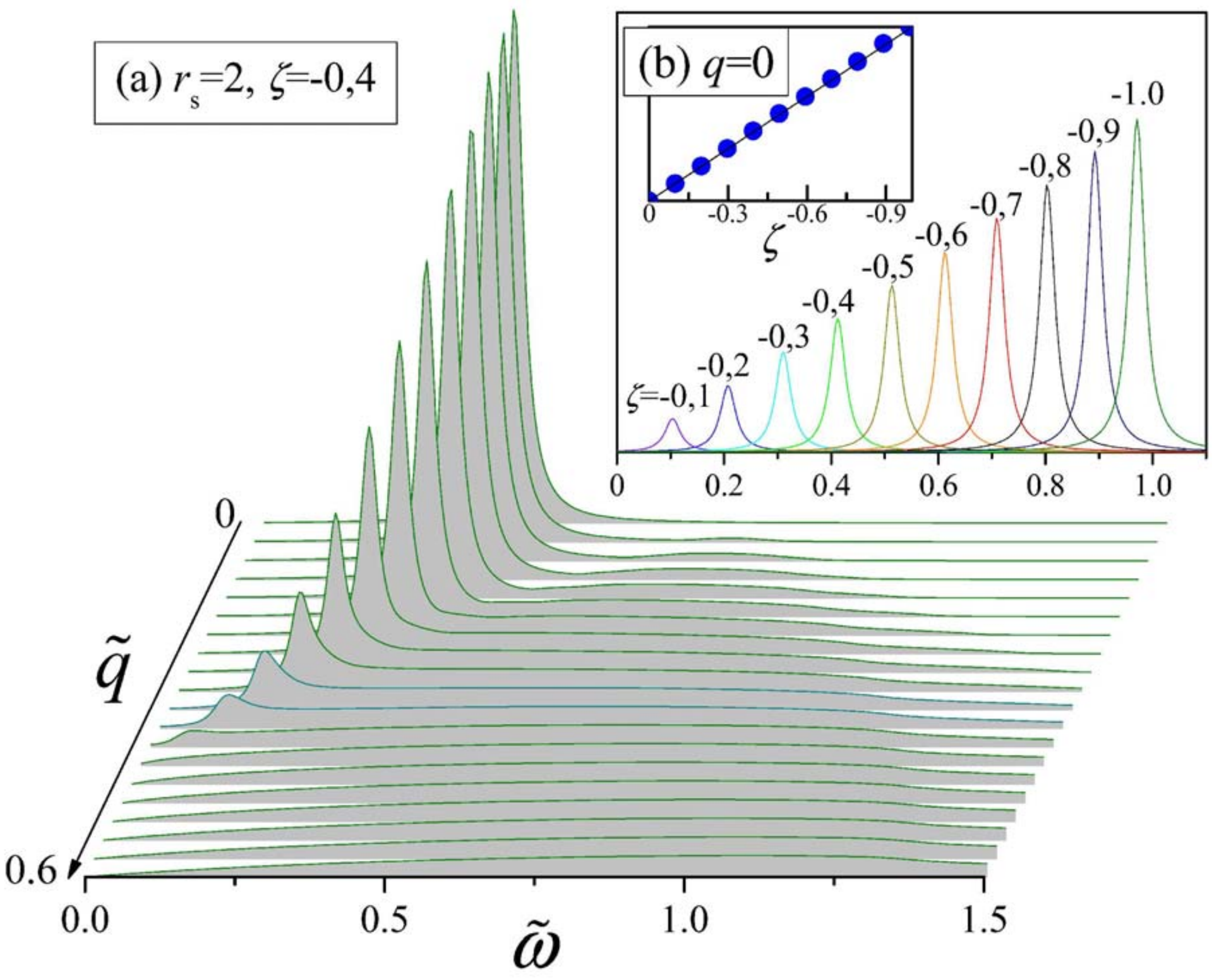

---